\documentclass[pra,amsmath,preprintnumbers,amssymb,aps,superscriptaddress,twocolumn]{revtex4-2}

\usepackage[utf8]{inputenc}
\usepackage[T1]{fontenc}
\usepackage[english]{babel}

\usepackage{hyperref}
\hypersetup{colorlinks=True,allcolors=blue}

 \usepackage{hhline}
 \usepackage[dvipsnames]{xcolor}
 \usepackage{dcolumn}
 \usepackage{graphicx}
 \usepackage{bm}
 \usepackage{amsmath}
 \usepackage{amsthm}
 \usepackage{amssymb}
 \usepackage{bbold}
 \usepackage{soul}
 \usepackage{filecontents}
 \usepackage{tikz-cd}
 \usepackage{float}
 \usepackage{placeins}
 \usepackage{amsthm}
 \usepackage[normalem]{ulem}
\usepackage{todonotes}
\usepackage{bbm}
\usepackage{booktabs}
\usepackage{appendix}

 \def\aop{a}
  \def\cop{c}
  
  \def\adag{a^{\dagger}}
  \def\cdag{c^{\dagger}}

\def\chaingain{\gamma_{\textsl{g}}}
\def\hoprate{\Gamma}
\def\chainloss{\kappa_{\ell}}
\def\cavloss{\kappa_c}

\newcommand{\change}[1]{\textcolor{black}{#1}}

\makeatletter
\let\ORIbbl@fixname\bbl@fixname
\def\bbl@fixname#1{
    {\ORIbbl@fixname#1}   {\edef\languagename{\@nameuse{languagealias@#1}}}%
}
\newcommand{\definelanguagealias}[2]{%
  \@namedef{languagealias@#1}{#2}%
}
\makeatother

\definelanguagealias{en}{english}
\definelanguagealias{eng}{english}

\begin{document}

 \title{Excitable quantum systems: the bosonic avalanche laser}

\author{Louis Garbe}
\email{Louis.garbe@wmi.badw.de}
\author{Peter Rabl}
 
\affiliation{Technical University of Munich, TUM School of Natural Sciences, 85748 Garching, Germany}
\affiliation{Walther-Meißner-Institut, Bayerische Akademie der Wissenschaften, 85748 Garching, Germany} 
\affiliation{Munich Center for Quantum Science and Technology (MCQST), 80799 Munich, Germany}

\begin{abstract}
We investigate the dynamics of a lasing system driven by a current of bosonic (quasi-)particles via a dissipative three-mode mixing process. A semi-classical analysis of this system predicts distinct dynamical regimes, where both the cavity mode and the gain medium can undergo lasing transitions. Of particular interest is an intermediate self-pulsing phase that exhibits the characteristics of an excitable system and converts random input signals into separated, quasi-periodic pulses at the output. By performing exact Monte-Carlo simulations, we extend this analysis into the quantum regime and show that despite being dominated by huge bosonic particle number fluctuations, this effect---reminiscent of coherence resonance--- survives even for rather low average photon numbers. Our system thus represents an intriguing model of an excitable quantum many-body system, with practical relevance for quantum detectors or autonomous quantum machines. As an illustration,  we discuss the realization of this system with superconducting quantum circuits and its application as a number-resolved avalanche detector for microwave photons. 
\end{abstract}

\maketitle

\section{Introduction} 
Excitable systems are a broad class of nonlinear dynamical systems that, in simple terms, can support propagating waves or other collective excitations, but cannot be re-excited until a certain amount of time has passed. Typical physical signatures of these systems are the effects of coherence resonance (CR) and stochastic resonance (SR), whereby a purely \textit{noisy} input can either generate a highly \textit{regular} response or amplify a weak periodic  signal~\cite{gang_stochastic_1993,pikovsky_coherence_1997,lindner_effects_2004,wellens_stochastic_2004}. Excitable systems are widely studied in various areas of classical physics, in particular in the context of chemical reactions and biological systems; however, comparably little is still known about such systems and their potential applications in the quantum regime. 
SR has been studied mainly in connection with thermally or externally driven transitions in bistable quantum systems~\cite{lofstedt_quantum_1994,grifoni_coherent_1996,grifoni_quantum_1996,whitney_temperature_2011,xie_interference_2018,hussein_spectral_2020,kato_quantum_2021,hanze_quantum_2021,li_stochastic_2024} and noise-assisted entanglement schemes~\cite{huelga_stochastic_2007,rivas_stochastic_2009}. CR has been observed in current oscillations in electronic lattices \cite{mompo_coherence_2018,bonilla_nonlinear_2024}, but still in a regime where the externally applied classical noise dominates over intrinsic quantum mechanical fluctuations. 

\begin{figure}
    \centering    
    \includegraphics[width=\columnwidth]{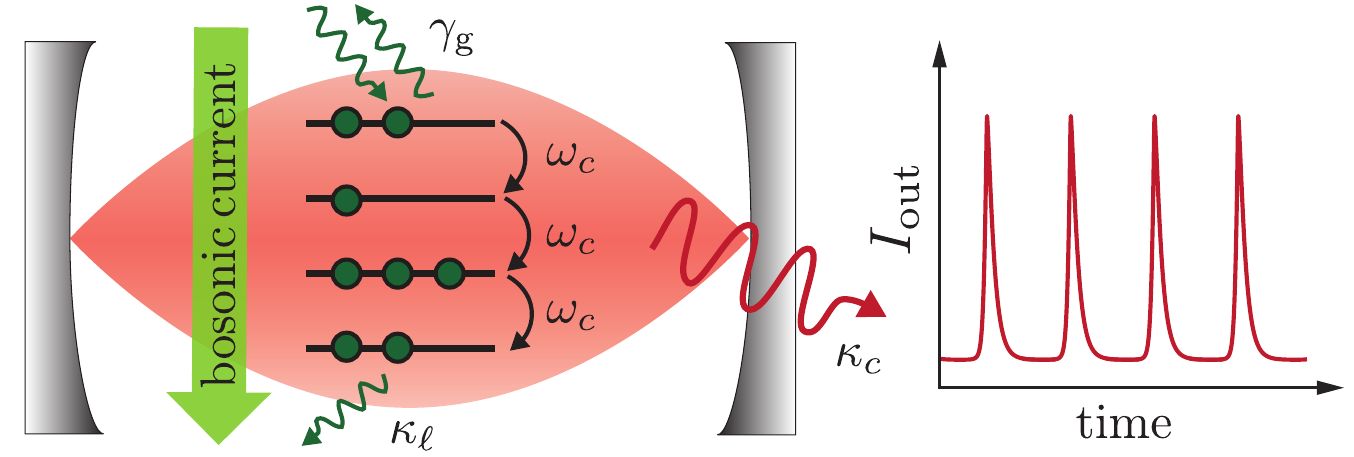}
    \caption{Sketch of a bosonic avalanche laser. Bosonic (quasi-) particles are injected randomly into the system with rate $\chaingain$ and transition down a ladder of $N\gg1$ equidistant energy levels by emitting a photon of frequency $\omega_c$ into the lasing cavity at each step. In turn, a high cavity occupation number accelerates the dissipative bosonic current, which, under certain conditions, can produce a semi-regular periodic signal at the output. See text for more details.}
    \label{fig:sketch}
\end{figure}

In this paper, we analyze the dynamics of a novel lasing system that is driven by a dissipative current of bosonic (quasi-) particles, as shown in Fig.~\ref{fig:sketch}. In this system, each of the injected bosons transitions between multiple intermediate energy levels, thereby emitting multiple photons into the cavity mode before leaving the system again. This process is similar to the concept of a bosonic cascade laser \cite{liew_proposal_2013,kaliteevskii_double-boson_2013,kaliteevski_single_2014,kavokin_bosonic_2016,liew_quantum_2016}, but here we assume that each transition is assisted by an auxiliary reservoir. This makes the transitions dissipative and, importantly, unidirectional. Under this condition, a non-trivial interplay between the bosonic current---which drives the cavity mode---and the cavity mode---which stimulates the current---emerges and gives rise to different lasing regimes. These include, in particular, a new dynamical self-pulsing phase, where, within a classical mean-field description, the system emits bursts of photons with a period that is a universal function of the underlying system parameters. Each burst is followed by a dark period with no emission, as is characteristic for excitable systems. 
With the help of exact Monte-Carlo simulations, we study the lasing system also for moderate and low excitation numbers, where the dynamics is dominated by the intrinsic shot-noise fluctuations that arise from the discreteness of the bosonic particle current. \change{From these simulations, we observe that even for very low average excitation numbers, signatures of self-pulsing survive, although the pulses are strongly affected by the noise. By analyzing in more detail the correlation spectrum of the output signal and the corresponding statistics of the emission burst, we show that the behavior of this lasing system is reminiscent of the phenomenon of CR: randomly injected bosons are converted into a semi-regular output signal. Specifically, we find that the coherence parameter, a measure of the regularity of the output pulses, exhibits a maximum as a function of pumping strength. Comparing different pumping mechanisms, we further show that random fluctuations in the injection can directly improve the regularity of the output, which is a key feature of excitable systems.
}

\change{
Deep into the quantum regime, we find that already a single boson injected into the system can produce an avalanche of signal photons---a mechanism that can be used as an amplifier for weak quantum signals. As a specific application, we propose an implementation of the described lasing system using superconducting quantum circuits and show that, under realistic experimental conditions, this device can serve, for example, as a number-resolved  detector for microwave photons. }

\section{A bosonic avalanche laser}

We consider the generic setup shown in Fig.~\ref{fig:sketch}, where a ladder of $N$ bosonic modes, the gain medium, is coupled to a lasing cavity via a dissipative three-wave mixing process. We model this system by a master equation of the form
\begin{equation} 
\begin{split}           
\dot \rho = \,& \hoprate \sum_{p=1}^{N-1} \mathcal{D}[\aop_p \adag_{p +1}\cdag] \rho  +\cavloss \mathcal{D}[\cop]\rho +\mathcal{L}_{\rm gain} \rho+ \mathcal{L}_{\rm loss} \rho,
\end{split} 
 \label{eq:Lindblad}
\end{equation}
where $\rho$ is the system density operator, $\mathcal{D}[C]\rho= C\rho C^\dag -(C^\dag C \rho- \rho C^\dag C)/2$, and  $\aop_p $  ($\aop^\dag_p $) and $\cop$ ($\cdag$) denote the bosonic annihilation (creation) operators of the ladder modes and the cavity mode, respectively. The first term in Eq.~\eqref{eq:Lindblad} describes bosons hopping from mode $p$ to the next lower mode $p+1$, while simultaneously emitting a photon into the cavity. This process is purely incoherent and irreversible, such that re-absorption events of the form $\mathcal{D}[\adag_p\aop_{p+1}\cop]$ do not occur. In Sec.~\ref{sec:Implementation} we discuss how, quite generically, such a dissipation process arises from a higher-order coupling to auxiliary bath modes. We also describe a more specific implementation of this process with superconducting circuits.

The photons in the cavity mode decay with a rate $\cavloss$. For a steady operation, gain bosons are injected continuously into mode $p=1$ with rate $\chaingain$ and leave the system through mode $p=N$ with rate $\chainloss$. We model those processes by the Lindbladians
\begin{equation}           
\mathcal{L}_{\rm loss} \rho= \chainloss \mathcal{D}[\aop_N] \rho 
\end{equation}
for loss and 
\begin{equation}           
\mathcal{L}_{\rm gain} \rho=  \chaingain (1-\zeta) \mathcal{D}[\aop_1]  \rho  +\chaingain \zeta  \mathcal{D}[\adag_1]   \rho, \label{eq:injection_lindbladian}
\end{equation}
for gain, respectively, 
where $\zeta\in[0,1]$. In the following, we focus primarily on the case $\zeta=1/2$, where the gain medium is coupled to an effective infinite-temperature reservoir. However, all effects discussed below are also observed for finite temperatures $(\zeta<1/2)$ or for models with pure gain $(\zeta=1)$. 

\subsection{Lasing regimes}
To describe the dynamics and steady state of this system, we start with a mean-field analysis, where we describe the cavity mode  by its classical amplitude, $\alpha_c=\langle\cop\rangle$, and we neglect correlations between the populations of the different ladder modes $n_p=\langle\adag_p\aop_p\rangle$. Under this approximation, we obtain the set of coupled equations,
 \begin{eqnarray} \nonumber
    \dot{n}_p&= & \left(1+\lvert\alpha_c\rvert^2\right)\left(J_{p-1,p}-J_{p,p+1}\right)+ \delta_{p,1}\chaingain  -\delta_{p,N}\chainloss n_p,\\
    \dot{\alpha_c}&= &\frac{1}{2}\left(J_{\rm cum}  -\cavloss\right)\alpha_c,
    \label{eq:MF_casc}
\end{eqnarray}
where $J_{p,p+1}=\hoprate n_p(1+n_{p+1})$ is the bosonic current between sites $p$ and $p+1$, $\delta_{i,j}$ is the Kronecker delta, and 
\begin{equation}
J_{\rm cum}=\sum_{p=1}^{N-1} J_{p,p+1}
\end{equation}
is the total cumulative current that is flowing through the device.  Eq.~\eqref{eq:MF_casc} clearly displays \textit{double stimulation}~\cite{kaliteevskii_double-boson_2013,kaliteevski_single_2014}, where the hopping of particles from site $p$ to $p+1$ is stimulated by both the cavity population $n_c=\lvert\alpha_c\rvert^2$ \textit{and} the number of bosons already present on the target site $n_{p+1}$.

\begin{figure}
      \centering      
      \includegraphics[width=\columnwidth]{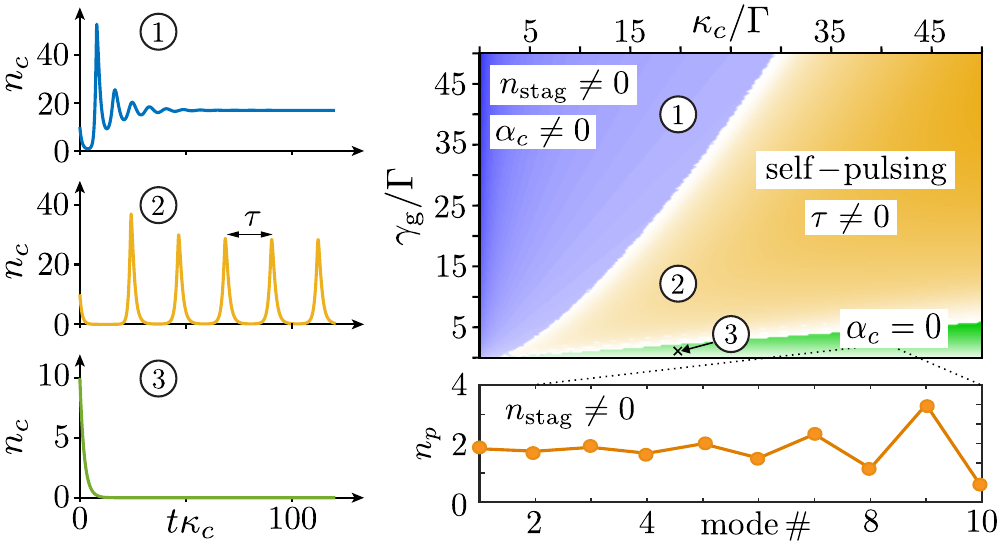}
      \caption{Mean-field phase diagram of the bosonic avalanche laser, as obtained from the solution of Eq.~\eqref{eq:MF_casc} (we used an initial seed amplitude $\alpha_c(t=0)=\sqrt{10}$ here, but the results are independent of the specific initial value). The different phases are distinguished by their characteristic dynamics, shown in the panels on the left for $\chaingain/\Gamma=2,12,40$ and $\kappa_c/\Gamma=20$, and by the order parameters $\alpha_c$ and $n_{\rm stag}$ in the steady state. Here, $|\alpha_c|>0$ indicates lasing of the cavity mode and $n_{\rm stag}\neq0$ condensation of the ladder modes~\cite{garbe_bosonic_2024} into a staggered density pattern,  which is shown in the panel below for $\chaingain/\Gamma=5$ and $\kappa_c/\Gamma=40$. In the self-pulsing phase (orange), no stationary state is reached.
      The other parameters used in these plots are $\kappa_\ell/\Gamma=10$, $\zeta=1/2$ and $N=10$. 
      }
      \label{fig:phasediag}
  \end{figure}

In Fig.~\ref{fig:phasediag} we show the transient dynamics and steady state of the mean-field equations for $N=10$ and varying pump and decay rates. In the limit $\cavloss/\chaingain \rightarrow 0$, the cumulative current, $J_{\rm cum}$, exceeds the cavity loss rate; the cavity mode is amplified and starts lasing. In turn, the amplified cavity mode accelerates the current through the ladder. This interplay results in an oscillatory initial dynamics, but the system quickly relaxes into a regular, i.e., stationary, lasing state of the cavity. In the opposite limit,  $\cavloss/\chaingain \rightarrow \infty$, the cavity is overdamped and stays de-excited. The remaining transport dynamics of the ladder modes in Eq.~\eqref{eq:MF_casc} then describes a so-called asymmetric simple inclusion process (ASIP), whose transient and stationary features have been analyzed in Refs.~\cite{garbe_bosonic_2024,Minoguchi2025}. Interestingly, this transport process can itself lead to lasing \textit{within the ladder}, where bosons condense into a few of the lowest modes with a characteristic zig-zag density profile~\cite{liew_quantum_2016,garbe_bosonic_2024}. This pattern, captured by the staggered population $n_{\rm stag}=\sum_p (-1)^p (n_p-n_1)$, is clearly visible in the density profile of the ladder modes shown in Fig.~\ref{fig:phasediag}. Note that a similar staggered configuration is also found in the cavity lasing phase, where the cavity mode can be replaced by a $c$-number that simply enhances the rate $\Gamma$ (see Appendix~\ref{sec:appendix_asip}). To avoid confusion, in the rest of the manuscript, we will use the terms "lasing" and "lasing mode" only to refer to the cavity mode.

In between these two limiting cases, i.e., for $\cavloss \sim \chaingain$,  the system does not reach a steady state. Instead, we observe a persistent emission of periodic photon bursts, which are separated by longer gaps during which the cavity mode remains de-excited. This behavior is self-sustained and characterized by a period $\tau\equiv \tau(\kappa_c,\chaingain,N,\hoprate)$, which is independent of the initial conditions in our mean-field simulations. 
Note that this behavior, as well as all the other features of the stationary phases discussed above, is robust with respect to intrinsic losses of the ladder modes, as long as those occur at a rate $\kappa_0\lesssim \Gamma$ (see Appendix~\ref{sec:appendix_loss}).

\subsection{Self-pulsing}

\begin{figure}
      \centering      \includegraphics[width=\columnwidth]{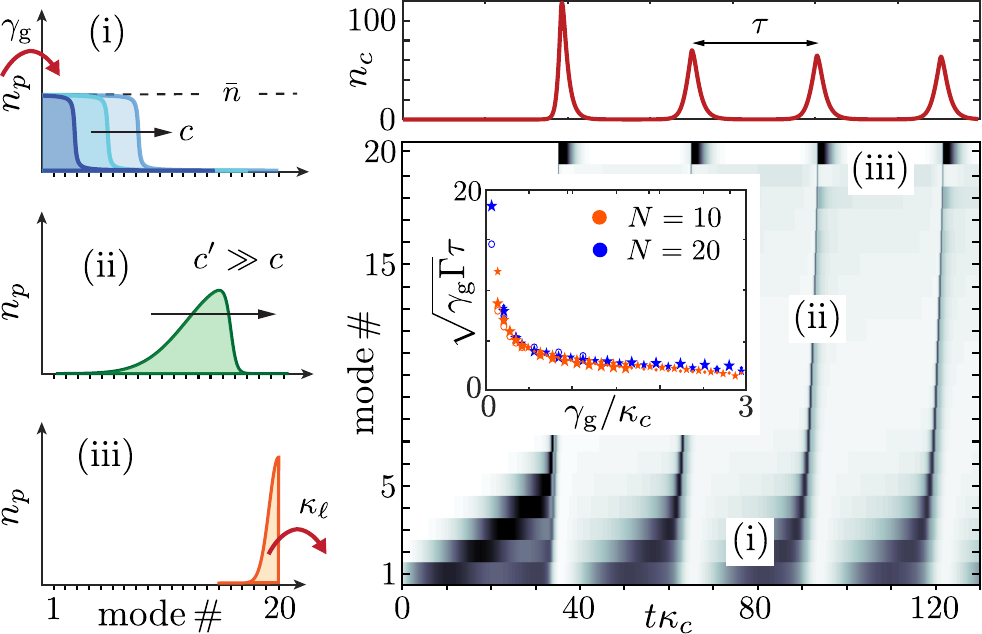}
      \caption{Origin of self-pulsing. The panel on the bottom right shows the time evolution of the mode occupation numbers $n_p$ over a few cycles for $\kappa_c/\Gamma=\kappa_\ell/\Gamma=20$ and $\chaingain/\Gamma=10$. The corresponding population of the cavity mode, $n_c$, is shown on top. In each cycle, the system evolves through three distinct phases, which are illustrated by the corresponding sketches on the left.  
      The inset shows the rescaled pulsing period $\tau$ obtained from mean-field simulations. The various symbols represent different combinations of $\cavloss/\kappa_\ell=5,25$ and $\hoprate/\kappa_\ell=0.05,5$, keeping $\kappa_\ell$ fixed. Additional parameter combinations are shown in Appendix~\ref{sec:appendix_period}.  The colors correspond to different values of  $N=10$ (red) and $N=20$ (blue). Upon rescaling, all curves collapse to the same universal behavior. For all plots, $\zeta=1/2$.  
      }
    \label{fig:packet_pulsing}
  \end{figure}

To understand the origin of the pulsing behavior, we plot in Fig.~\ref{fig:packet_pulsing} the individual populations $n_p$ over several periods. The dynamics can be decomposed into three stages, sketched on the left-hand side of Fig.~\ref{fig:packet_pulsing}. (i) Starting from an empty system, bosons are injected into mode $p=1$ and a density wave starts to flow through the ladder with a speed $c\approx (1+\bar{n})\hoprate$, enhanced by the quasi-stationary occupation $\bar{n}$ of the first few modes. Assuming this occupation to be roughly constant throughout the packet, it can be determined by the left boundary condition, which reads $\hoprate \bar{n}(1+\bar{n})=\chaingain$. 
During this initial phase, the cumulative current $J_{\rm cum} \sim \chaingain ct$ increases approximately linearly with time, and eventually exceeds the cavity loss, $\cavloss$. (ii) Beyond this point, the lasing mode is amplified; once the cavity population has become significantly larger than unity, the bosonic density wave is accelerated accordingly to a speed $c' \approx (1+|\alpha_c|^2) c \gg c$. (iii) This speed-up empties the ladder almost instantaneously, after which the cavity mode decays, and the cycle starts anew. 

Given that the dynamics of the lasing burst itself is quite fast, we expect the period $\tau$ of the total process to be mainly determined by the initial build-up stage. We can express this condition approximately as $J_{\rm cum}(\tau)\approx \cavloss$, with  $J_{\rm cum}(\tau)\approx \chaingain (1+\bar{n})\hoprate\tau$.  For $\bar{n}\gg1$, we then obtain a scaling $\sqrt{\chaingain \hoprate}\tau \sim (\cavloss/\chaingain)$, roughly independent of the number of ladder modes. In the inset of Fig.~\ref{fig:packet_pulsing}, we plot the rescaled mean-field period $\sqrt{\chaingain \hoprate}\tau$ for various different ratios between the parameters $\hoprate$, $\chaingain$, and $\cavloss$, and two different values of $N$. We observe indeed a collapse of all results onto a single universal function, although the precise dependence deviates from the estimate above, which oversimplifies the actual evolution of the gain current and the cavity mode during the build-up stage. A more detailed description of the pulse-generating mechanism, and of the scaling of the pulsing period, can be found in  Appendix~\ref{sec:appendix_period}.

\section{Quantum stochastic dynamics}
\label{sec:QSD}
Given the ability of our lasing system to support self-sustained excitations at the mean-field level, the question remains whether or not such a behavior prevails in the presence of noise and when the full many-body dynamics is taken into account. In the current setting, fluctuations arise intrinsically from the discrete nature of the bosonic current, which generates \emph{bosonic shot-noise} at injection and during transport~\cite{Minoguchi2025}. \change{In the quantum regime with low photon numbers, these fluctuations are expected to  wash out the pulsing cycles. In excitable systems, however, the mechanisms of SR and CR can also induce the opposite effect and convert the noisy input into regular output signals. }

To investigate the role of these competing effects on the overall behavior of our lasing system, we use a stochastic unraveling of Eq.~\eqref{eq:Lindblad} together with Monte-Carlo techniques to perform exact simulations of the population dynamics of the gain and the cavity modes (see Appendix~\ref{sec:appendix_numerics} for the details of the numerical method). Note that the noise process used in those simulations depends on the pumping mechanism defined by $\mathcal{L}_{\rm gain}$ in Eq.~\eqref{eq:injection_lindbladian}. For now, we focus on the infinite-temperature limit $\zeta=1/2$, for which---for a given average pump rate---fluctuations are maximal. \change{In our simulation, this process is implemented by randomly injecting particles in the first chain site at a rate $\gamma_{\rm inj}=\gamma_\textsl{g}(1+n_1)$, and extracting them at a rate $\gamma_{\rm extr}=\gamma_\textsl{g}n_1$, with $n_1$ the number of particles already present on the first site. The average gain rate is then given by $\gamma_{\rm inj}-\gamma_{\rm ext}=\gamma_\textsl{g}$.}  \change{We return to a discussion about other types of injection processes in Sec.~\ref{sec:discussion_pumping} below.} 

\change{In Fig.~\ref{fig:SCR}(a), we display sample trajectories of the cavity population in the three phases predicted by our mean-field analysis. All cases show the dominating effect of quantum fluctuations, and there is no longer a clear separation between the different lasing regimes. However, we still observe the emission of semi-regular photon bursts. To quantify this behavior, we follow two different approaches and analyze, on the one hand, the noise spectrum of the output field, 
and on the other hand, the inter-spikes interval statistics.}
\begin{figure}
    \centering    
     \includegraphics[width=\columnwidth]{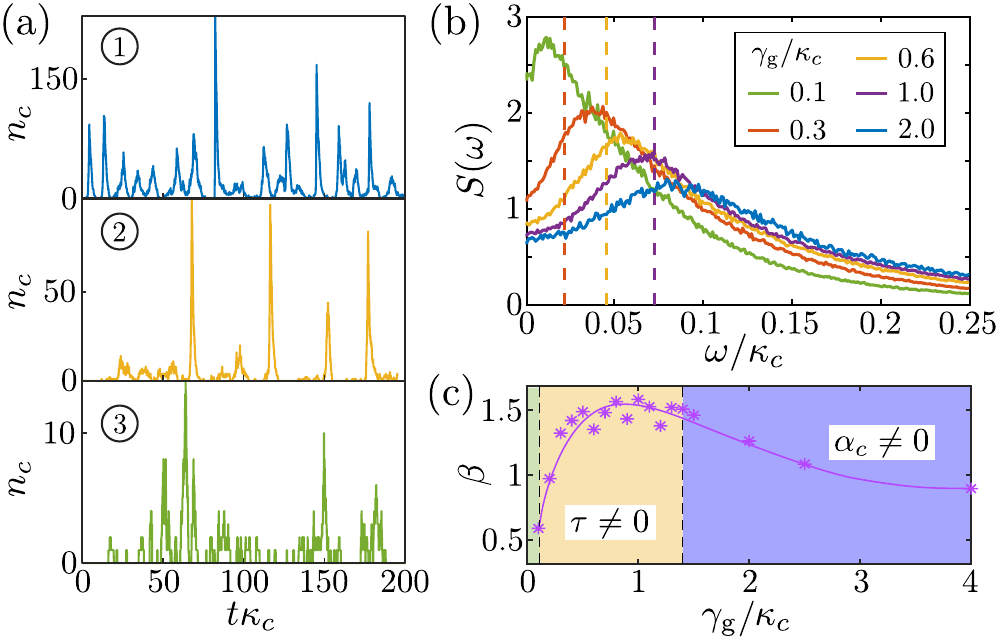}
    \caption{(a) Typical stochastic trajectories of the cavity photon number $n_c$ (same parameters as the mean-field results shown in Fig.~\ref{fig:phasediag}).
    (b) Plot of the noise spectrum for different values of $\chaingain$, which characterizes both the average pumping rate and the strength of the intrinsic bosonic shot-noise.  
    (c) The coherence parameter $\beta$ defined in Eq.~\eqref{eq:beta} is plotted as a function of $\chaingain/\cavloss$, where the stars represent the numerically evaluated values and the solid line is a guide to the eye. A maximum of $\beta$ is found around $\chaingain/\cavloss\approx 1$. For all plots, we have assumed $N=10$, $\kappa_\ell/\Gamma=20$, and $\zeta=1/2$.}   
    \label{fig:SCR}
\end{figure}

\subsection{Noise spectrum and coherence parameter} 
First, we evaluate the correlation function
\begin{align}
    C(s)&=\frac{1}{T}\int_0^{T} n_c(t)n_c(t+s) dt-\left(\frac{1}{T}\int_0^{T} \, n_c(t) dt\right)^2
\end{align}
of each trajectory for a sufficiently large time interval $T$ and define the normalized noise spectrum as
\begin{align}
   S(\omega)= \kappa_c \int ds\,   \left\langle\frac{C(s)}{C(0)}\right\rangle e^{i\omega s},
\end{align}
where $ \langle \bullet\rangle$ denotes the trajectory average. The shape of the resulting noise spectra is plotted in Fig.~\ref{fig:SCR}(b) for different $\chaingain$, which quantifies the strength of the driving process and of the injected noise. We see that for all parameters, the spectrum reaches a maximum, $S(\omega_{\rm max})$, at a nonvanishing frequency $\omega_{\rm max}>0$, which indicates a semi-regular behavior. Deep within the self-pulsing phase, the peak frequency $\omega_{\rm max}\approx 2\pi/\tau$ is roughly consistent with the mean-field prediction for the period $\tau$, although we do not find a tight correspondence in general.

In the literature on excitable systems, it is common to introduce the so-called coherence parameter~\cite{gang_stochastic_1993,kato_quantum_2021}
\begin{equation}\label{eq:beta}
\beta=\frac{\omega_{\rm max}}{\Delta\omega} S(\omega_{\rm max}), 
\end{equation}
where $\Delta \omega$ is the half-width at half-maximum of the spectral peak. In essence, the parameter $\beta$ quantifies the regularity of the signal produced; a characteristic feature of excitable systems is the existence of a maximum of the coherence parameter as a function of the applied noise strength. For the current system, $\beta$ is plotted in Fig.~\ref{fig:SCR}(c) for varying $\chaingain$, from which we make two important observations. First, the coherence parameter changes smoothly across the phase boundaries shown in Fig.~\ref{fig:phasediag}, which demonstrates that nonlinear features of classical excitable systems do not necessarily translate into the quantum regime. \change{However, $\beta$ still exhibits the characteristic maximum, which is reminiscent of the behavior found in excitable systems exhibiting CR. We will discuss this connection further below. } 

\change{
The results presented in Fig.~\ref{fig:SCR}, which are obtained for the limit of an infinite-temperature bosonic reservoir with $\zeta=1/2$, show that even this maximally noisy input is converted into a semi-regular output. In Appendix~\ref{sec:pumpmech_appendix}, we also present simulation results for a pure gain process with $\zeta=1$, for which we observe an even more pronounced maximum of the coherence parameter.}

\subsection{Inter-spike interval statistics}
\label{sec:burst_stat}
\begin{figure}
    \centering    \includegraphics[width=\linewidth]{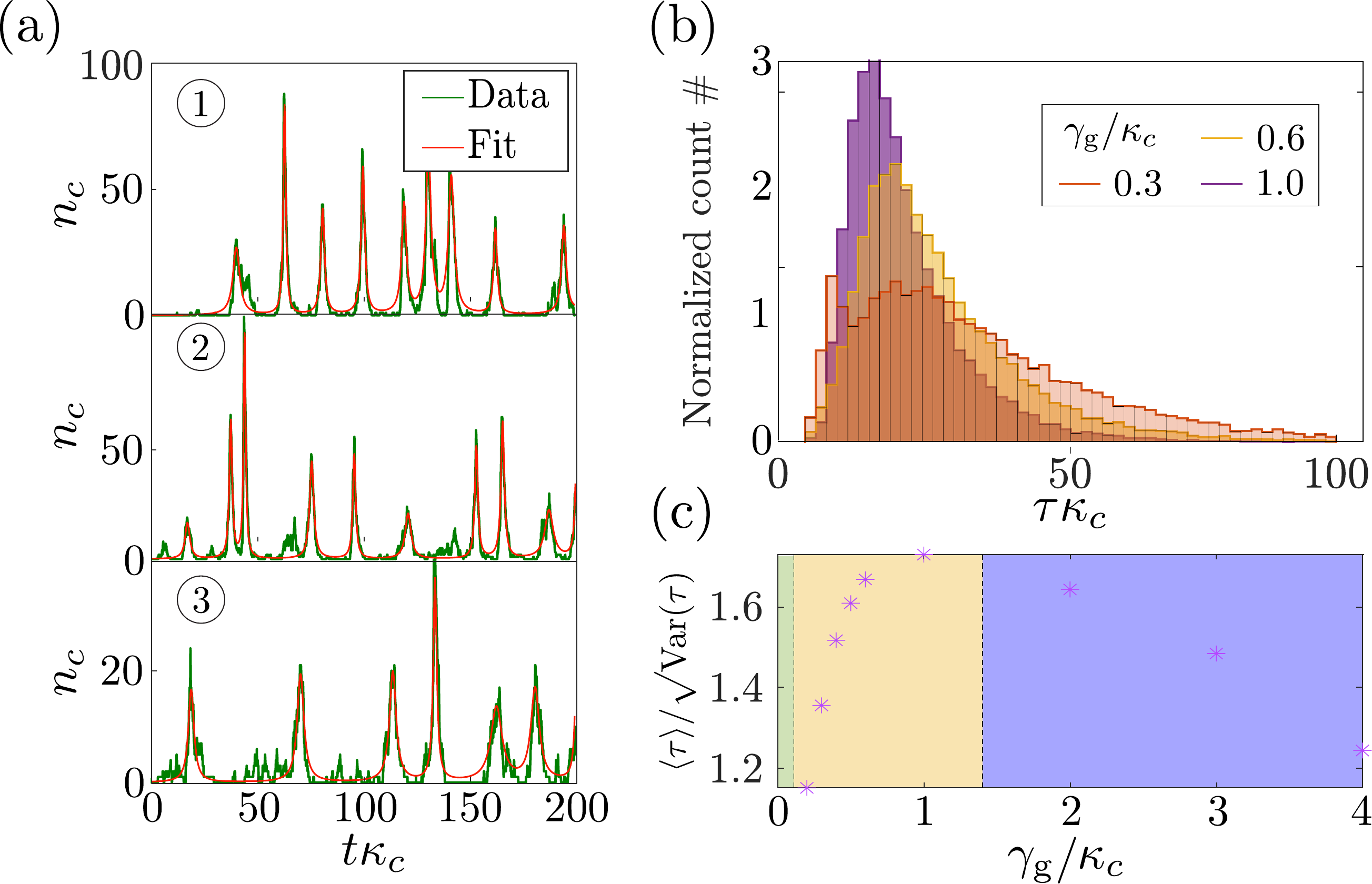}
    \caption{\change{Inter-spike interval statistics. (a) Illustration of the fitting procedure used to identify pulses in the output of the lasing cavity. The actual trajectories obtained from Monte-Carlo simulations (blue lines) are overlapped with the fitted Lorentzian curves (red lines), whose maxima lie at times $t_{\rm b}^{(i)}$. For these examples, we have assumed the ratios $\gamma_{\rm g}/\kappa_c=0.3,0.6,1$ (from top to bottom). (b) Histogram of the intervals $\tau_i$ between successive bursts, for the same three $\gamma_{\rm g}/\kappa_c$ values.
    (c) Plot of the ratio between the mean burst interval and its standard deviation as a function of the pump strength. All other parameters in the plots are the same as in Fig.~\ref{fig:SCR}.}
    }   \label{fig:SCR_intervalstat}
\end{figure}

\change{To consolidate these findings, we also conducted a direct analysis of the inter-burst intervals. Contrary to other excitable systems \cite{bonilla_nonlinear_2024}, the pulses in our setup present fairly large fluctuations in both height and emission time. Thus, to identify a burst, we only take into account signals that exceed a certain threshold (here set at twice the time-average of the signal). We then fit the part of each trajectory that lies above this threshold by Lorentzian functions, in order to remove artifacts coming from short-time oscillations and multiple subpeaks. This fitting procedure is illustrated for a few sample trajectories in Fig.~\ref{fig:SCR_intervalstat}(a).  From the centers of the Lorentzians, we obtain the set of burst times $t_{\rm b}^{(i)}$, from which we sample the distribution of interval spacings $\tau_i=t_{\rm b}^{(i)}-t_{\rm b}^{(i-1)}$.} 

\change{The resulting distributions for the interval spacings are presented in  Fig.~\ref{fig:SCR_intervalstat}(b) for different values of $\gamma_{\rm g}$. As we increase the pumping strength, the interval distribution shifts towards smaller values, while its variance tends to decrease. This is in agreement with the analysis in frequency space, which showed noise spectra acquiring a larger mean and variance for increasing pumping strength. Furthermore, the maximum of the distribution, although it deviates from the mean-field prediction, largely coincides with the corresponding maximum in the noise spectrum. To quantify the regularity of the distribution, we plot in Fig.~\ref{fig:SCR_intervalstat}(c) the ratio between the mean interval spacing and its standard deviation, $\langle \tau\rangle/\sqrt{\langle \tau^2\rangle-\langle \tau\rangle^2}$. This quantity exhibits a maximum around $\gamma_\textsl{g}/\kappa_c=1$, again in line with the behavior in the frequency domain. }

\subsection{Role of the noise}\label{sec:discussion_pumping}

\change{
In the numerical simulations presented above, we have observed a maximum for the coherence parameters in both frequency and time space.
Both quantities indicate a certain amount of regularity of the emitted pulses, in spite of an overall strongly fluctuating behavior. These maxima are reached as a function of the gain rate $\gamma_{\rm g}$, which, in view of the shot noise of the injected particle current, is also a measure for the amount of noise that is injected into the system. The observed behavior is thus reminiscent of CR in classical excitable systems, where similar maxima in the coherence parameter occur. CR, however, is usually associated with a quasi-regular response of a noise-driven system, which shows no particular dynamics in the absence of the noise. Here, instead, self-oscillations may already occur at the mean-field level.}

\change{Leaving aside the precise correspondence between the observed quasi-regular pulses and SR and CR in the conventional sense, it would be interesting to see more directly the influence of noise on the production of a semi-regular output. In the current model, this influence is somewhat hidden by the fact that the ladder is driven by an average, i.e., deterministic current, with shot-noise fluctuations on top. To separate those two contributions, we performed additional numerical simulations, where we replaced the gain process in Eq.~\eqref{eq:injection_lindbladian} by}
\begin{equation}
\mathcal{L}_{\rm gain} \,\, \rightarrow  \,\, x \mathcal{L}_{\rm gain}+ y \mathcal{L}_{\rm Poisson}+ (1-x-y) \mathcal{L}_{\rm reg}, 
\label{eq:compadet_process}
\end{equation}
\change{where $0\leq x+y\leq 1$. Here, $\mathcal{L}_{\rm reg}$ and $\mathcal{L}_{\rm Poisson}$ both describe incoherent processes in which bosons are injected one by one into mode $p=1$, independently of the prior occupation of the mode. This is to be contrasted with $\mathcal{L}_{\rm gain}$, in which driving is achieved by a balance of gain and loss (see discussion at the beginning of Sec.~\ref{sec:QSD}). For $\mathcal{L}_{\rm reg}$, the injection occurs at quasi-regular times, spaced by $\gamma_\textsl{g}^{-1}$ \footnote{More precisely, the probability that an injection takes place in the interval $[\tau, \tau+d\tau]$ obeys a Gaussian distribution: $p_{e}(\tau,\tau+d\tau)=\frac{2}{\Delta\sqrt{\pi}}e^{-\left(\frac{\tau-\tau_0}{\Delta}\right)^2} dt$. In the limit $\Delta\rightarrow0$, particles are injected exactly at times $\tau_n=n\tau_0$, thereby realizing a deterministic injection at a rate $1/\tau_0=\gamma_\textsl{g}$. In practice, we found that values of $\Delta \leq \gamma_\textsl{g}$ were enough to achieve convergence.}. For $\mathcal{L}_{\rm gain}$, by contrast, these intervals follow a random Poisson distribution, again with average $\gamma_\textsl{g}^{-1}$\footnote{At the Lindbladian level, this would be realized by a jump operator:
$\mathcal{L}_\text{gain}=\mathcal{D}\left[\frac{1}{\sqrt{1+a^+a}}a^+\right]$}.}
\change{This means that, for $x=1,y=0$ or $x=0,y=1$, we recover a noisy drive, respectively through an infinite-temperature reservoir or a particle-by-particle Poisson process; whereas, for $x=y=0$, the injection becomes quasi-deterministic. By varying $x$ and $y$, this model allows us to tune the level of randomness in the gain process, while keeping the same average injection rate.}
\change{Note that the process $\mathcal{L}_{\rm Poisson}$ may actually be quite difficult to engineer experimentally, but its theoretical appeal is to offer a direct counterpart to the deterministic process; since these two processes differ only by the injection interval distribution, any difference between them can be fully attributed to the stochastic nature of the drive.}

\begin{figure}
    \centering
    \includegraphics[width=\linewidth]{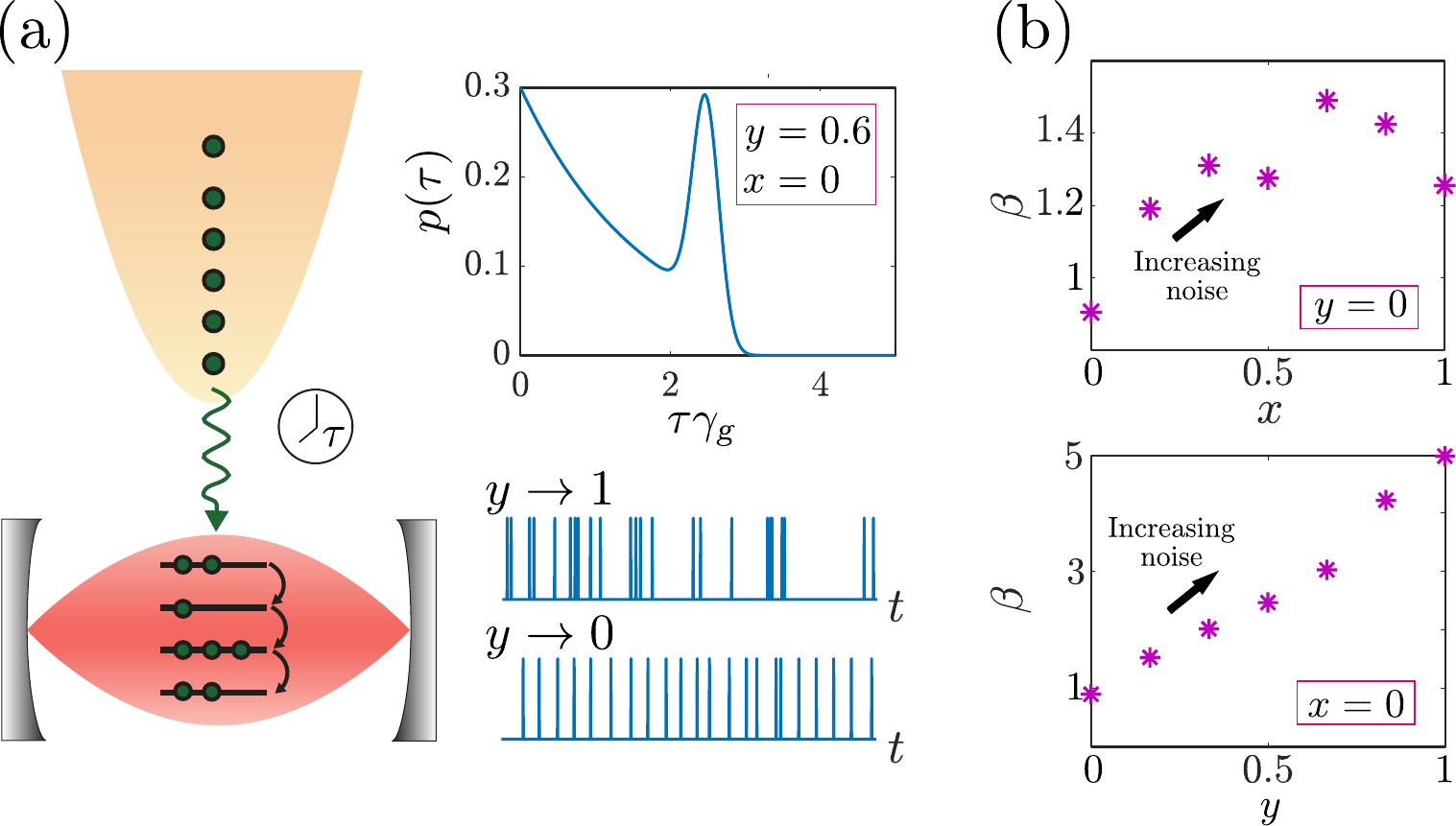}
    \caption{\change{(a) Pumping scheme of Eq.~\eqref{eq:compadet_process}, for $x=0$. Particles are injected one by one at random intervals $\tau$, whose distribution $p(\tau)$ is the sum of a stochastic and a quasi-deterministic process. The limits $y\rightarrow 1$ and $y\rightarrow 0$ give respectively fully random and fully periodic injection events. For $x\neq 0$, we have instead a balance between injection and extraction processes. 
    (b) Coherence parameter versus stochasticity of the drive, for a constant total rate $\gamma_\textsl{g}=0.6\kappa_c$. The quasi-deterministic process $\mathcal{L}_{\rm reg}$ is compared to an infinite-temperature bath (top) or a pure Poisson process (bottom). In both cases, the added stochasticity contributes positively to the regularity of the output.}}
    \label{fig:SCR_stochvsdet}
\end{figure}

\change{In Fig.~\ref{fig:SCR_stochvsdet}, we show that a larger $x$ or $y$ (therefore, a noisier process) can  significantly improve the coherence of the output. For the parameters displayed here, there is even an optimal value of $x$ for which the coherence is maximized, which is again suggestive of CR.
By isolating the contribution of random fluctuations in the drive, this analysis thus proves that noise does help in creating a more coherent output, showing how this key feature of excited systems is carried out in the quantum regime.}

\change{In general, the problem of separating the stochastic and deterministic contributions in the drive is not straightforward when dealing with weakly-driven systems in the quantum regime: we discuss this issue in more details in Appendix~\ref{sec:appendix_discussion}. Beyond the specific solution we have adopted to characterize the system at hand, a more systematic approach may be needed to study further examples of quantum excitable systems. We leave this point to future investigations.} 

\section{Applications: An avalanche detector for microwave photons}\label{sec:Implementation}
Beyond its fundamental interest, the ability to convert even a few injected gain bosons into an avalanche of photons at the cavity output makes this mechanism very promising for quantum sensing and amplification applications. As a specific use-case, we discuss in this section the realization of our model in Eq.~\eqref{eq:Lindblad} with superconducting circuits, and illustrate its use for microwave photon detection. 

\subsection{Circuit QED implementation: general approach}
\begin{figure}
    \centering    \includegraphics[width=\columnwidth]{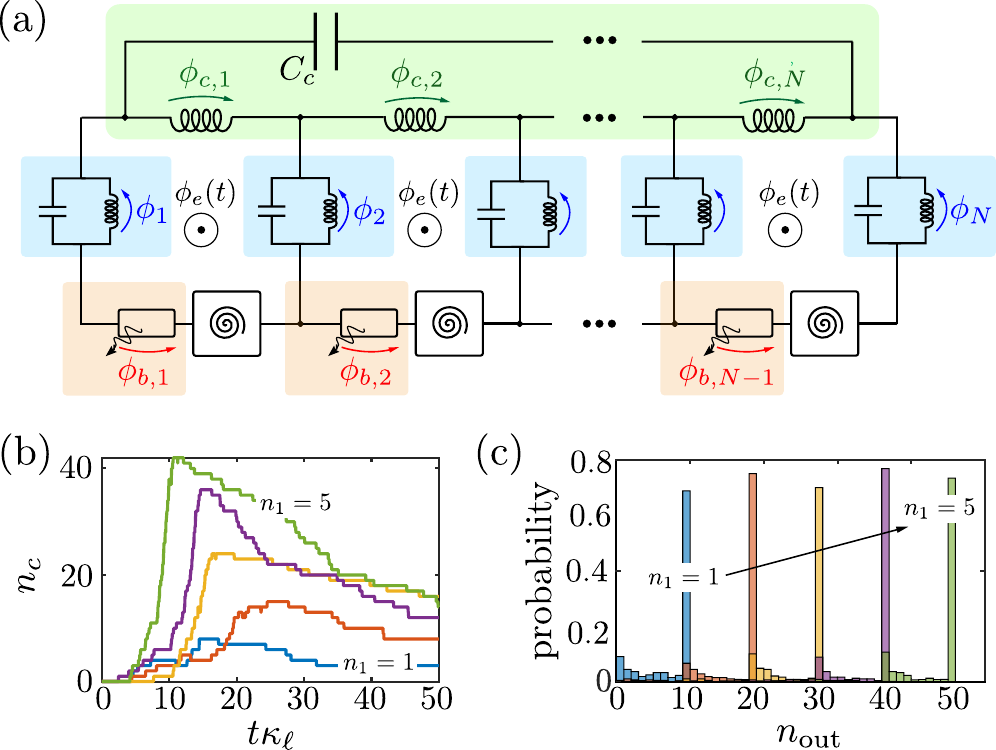}
    \caption{(a) Sketch of a superconducting circuit layout for realizing a bosonic avalanche laser. A set of $LC$ resonators representing both the common cavity mode (green) and the ladder modes (blue) are coupled via nonlinear SNAIL-type~\cite{Frattini2017} couplers (spiral) to an additional set of strongly damped waste modes (orange). As discussed in Appendix~\ref{app:Implementation}, the coupling can be modulated by external fluxes $\phi_e(t)$ to achieve a resonant four-mode interaction of the form given in Eq.~\eqref{eq:Vfourphoton}. After eliminating the waste modes, we obtain the dissipative three-mode hopping in Eq.~\eqref{eq:Lindblad}.   
(b) Sample trajectories showing the cavity population $n_c$ over time for different initial populations $n_1$ of the first ladder mode. For the same conditions, (c) shows the corresponding distribution of the integrated output signal of the cavity mode, $n_{\rm out}$, for $500$ trajectories. For these plots, we have assumed $N=10$, $\kappa_\ell=10\Gamma$, $\kappa_c=0.2\Gamma$, and also that each ladder mode decays with a rate $\kappa_0=0.2\Gamma$.
}
    \label{fig:circuit}
\end{figure}

To realize the cascaded lasing process in the microwave regime, we consider a superconducting circuit as outlined in Fig.~\ref{fig:circuit}(a). In this circuit, the extended cavity mode (shaded in green) as well as the individual ladder modes (shaded in blue) are represented by quantized $LC$ resonators with frequencies $\omega_c$ and $\omega_p$ in the microwave regime. To induce the targeted nonlinear dissipation processes, each pair of neighboring ladder modes is coupled to the cavity and an additional dissipative waste mode (shaded in orange), with frequency $\omega_b$ and annihilation operator $b_p$, through a nonlinear Josephson coupler. This coupling element generates interaction terms at all orders between the various modes. Here we are interested in fourth-order contributions of the form
\begin{equation}\label{eq:Vfourphoton}
     V_p\simeq g \left( \aop_p\adag_{p+1} b^\dag_{p} \cop^\dag  + {\rm H.c.}\right),
\end{equation}
which can be selected among other fourth-order and lower-order terms by driving the system with an externally applied magnetic flux at a frequency $\omega_e=\omega_p-\omega_{p+1}+\omega_{b}+ \omega_c$. This coupling generates the process $ \aop_p\adag_{p+1}\cop^\dag$ that we are interested in, plus the creation of an additional excitation in the waste mode.
Photons created in this mode quickly decay with a rate $\kappa_b\gg g$, and can thus be adiabatically eliminated to obtain the incoherent hopping terms in Eq.~\eqref{eq:Lindblad}, with a rate $\Gamma\simeq 4g^2/\kappa_b$. The full model is then achieved by repeating this unit cell in a chain configuration. 

Following this general approach, we present in Appendix~\ref{app:Implementation} a more detailed discussion of the individual circuit elements and the main assumptions that are necessary to arrive at the process given in Eq.~\eqref{eq:Vfourphoton} and the resulting master equation in Eq.~\eqref{eq:Lindblad}. This analysis shows that, for state-of-the-art experimental parameters, the realization of a bosonic avalanche laser with $N=5-10$ ladder modes and a hopping rate of $\Gamma/(2\pi)\approx 100$ kHz can be achieved. This rate exceeds the intrinsic loss rate $\kappa_0/(2\pi)= 10-100$ kHz of high-Q microwave resonators~\cite{Frunzio2005,Reagor2013,Somoroff2023}, such that individual photons can still hop through the whole ladder before they decay.

\subsection{Microwave Photon Detection}

Based on the parameters estimated for the circuit QED implementation above,  we simulate the cavity population under the assumption that the first ladder mode is initialized with a well-defined number of photons, $n_1=1,\dots,5$. In Fig.~\ref{fig:circuit}(b), we first show the successive evolution of the cavity population for individual trajectories of $n_c(t)$. 
We see that already a single photon can trigger a large excitation of the lasing cavity mode, despite significant losses in each of the ladder modes. As the initial number of photons increases, the build-up of this excitation is both accelerated and amplified, consistent with the mechanism described in Fig.~\ref{fig:packet_pulsing} for large photon numbers. For the same parameters, Fig.~\ref{fig:circuit}(c) shows a histogram of the integrated output signal, 
\begin{equation}
n_{\rm out}= \int_0^T dt \, I_c(t),
\end{equation}
where $I_c(t)=\kappa_c n_c(t)$ is the photon current emitted by the cavity. We see that different initial photon-number states give rise to output signals that are well separated, approximately by a photon number corresponding to the number of ladder modes. The residual overlap between neighboring distributions can be attributed to the loss of a photon in the initial mode, $p=1$, before the accelerated cascade process has started. Losses at a later stage are less relevant. Therefore, when combined with a regular microwave amplifier capable of resolving those amplified wavepackets, this device realizes a number-resolved detector for microwave photons.

\section{Conclusion}
In summary, we have analyzed the behavior of a bosonic avalanche laser, in which a dissipative three-mode mixing process converts a bosonic input current into an amplified signal at the cavity output. We have shown that in this system, mutually stimulated processes can lead to a self-pulsing phase at the mean-field level, which survives as a train of noisy but semi-regular output bursts even deep in the quantum regime, where the dynamics is dominated by bosonic shot-noise. Interestingly, the regularity of these bursts can be enhanced by increasing the amount of noise, and we observed a maximum of the coherence parameter, reminiscent of CR in conventional excitable systems.   
This conversion of quantum fluctuations into an amplified, semi-regular signal can potentially be used in quantum sensing applications, and we described the implementation of a number-resolved microwave photon detector as a specific example. The underlying mechanism and the described behavior can also be of relevance for the development of autonomous quantum engines and clocks~\cite{Mari2015,SerraGarica2016,Carollo2020,Guzman2024}, where the interplay between periodic motion and quantum fluctuations is at the heart of the subject and still not well understood.

\acknowledgments
We thank Mathias Marconi, Lukas Schamriss and Kirill Fedorov for helpful discussions. This research is part of the Munich Quantum Valley, which is supported by the Bavarian state government with funds from the Hightech Agenda Bayern Plus.


\appendix

\section{ASIP dynamics and population profile} \label{sec:appendix_asip}

In the limit of large $\kappa_c$, the cavity amplitude $\alpha_c$ can be set to zero, and the dynamics described by \eqref{eq:MF_casc} becomes a process for the ladder of modes only, described by the so-called ASIP model \cite{garbe_bosonic_2024}. 
When the population distribution is smooth, we may approximate the discrete populations $n_p$ by a continuous field $n(x)$, that obeys the partial differential equation:
\begin{equation}
\partial_t n=\Gamma\left[(1+2n)\partial_x n+\frac{1}{2}\partial^2_x n\right]
\end{equation}
This dynamics is characterized by a competition between ordinary diffusion and \textit{non-linear} advective transport, induced by the bosonic statistics of the particles.
In the presence of boundary drive and dissipation, and in the steady-state regime, this competition generates a pattern of oscillating population, whose amplitude grows near-exponentially as one goes towards the last site, and which we dubbed \textit{bosonic skin effect} \cite{garbe_bosonic_2024}.\\

Quantitatively, one may rewrite the steady-state condition as $J_{p,p+1}=J_{ss}\hspace{5pt} \forall p$, i.e., the steady-state imposes a homogeneous, constant current, which can be obtained from the boundary condition. In the case 
$\zeta=1/2$, which we consider in the main text, we get simply $J_{ss}=\gamma_\textsl{g}$. The steady-state condition then becomes a recurrence relation for the population: $n_p=\frac{J/\Gamma}{1+n_{p+1}}$, from which we can derive the alternating pattern. Near the left boundary, in which the oscillation amplitude vanishes, we have $n_2\sim n_1$, from which we can infer $n_1\sim\sqrt{\gamma_\textsl{g}/\Gamma}$. The population on the last site can also be obtained through the boundary condition, giving us access to the full oscillation profile (see \cite{garbe_bosonic_2024} for the full analytical expression).\\

Perturbations around this steady-state, in the form of small population fluctuations $\epsilon_p=n_p-n_1$, obey a linearized wave equation $\partial_t \epsilon=c\partial_x\epsilon+\frac{\Gamma}{2}\partial^2_x\epsilon$; these fluctuations therefore propagate with an effective speed
\begin{equation}
c=\Gamma(1+2n_1)\sim 2\sqrt{\Gamma \chaingain}.
\end{equation}
The presence of the $n_1$ factor indicates that the effective speed is renormalized by the average filling of the ladder, which is again a consequence of the stimulation of the transport by other bosons in the ladder. This quantity provides the typical time-scale at which perturbations propagate along the ladder and eventually subside; in the pulsing phase, it provides the scale to compare the oscillation period to (see section~\ref{sec:appendix_period}).\\

Finally, in the lasing phase, the cavity population will settle to a non-zero value. The dynamics in the chain will once more be described by the ASIP model, with an effective hopping rate $\Gamma(1+\lvert\alpha_c\rvert^2)$ rescaled by the cavity population. The steady-state will once more display a modified staggered pattern determined by this effective hopping rate.

\section{Losses in the ladder}
\label{sec:appendix_loss}
\begin{figure}
    \centering 
    \includegraphics[width=\linewidth]{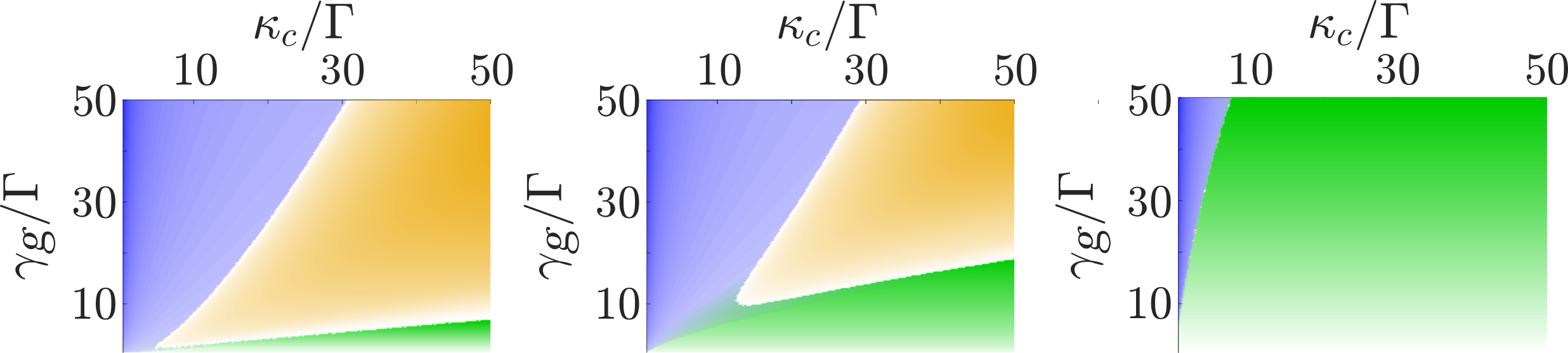}
    \caption{Phase diagram in the presence of losses on each ladder site: $\kappa_0/\hoprate=0.1$ (left), $1$ (center), $10$ (right). Other parameters are $\kappa_\ell/\Gamma=10$, $\zeta=1/2$, and $N=10$. 
    The main features of the three phases remain essentially unchanged even for $\kappa_0\sim \Gamma$.}
    \label{fig:PD_withchainloss}
\end{figure}

An implementation of the model described here will necessarily involve dissipation channels beyond the one we described. As we discussed in the End Matter, channels inducing only loss of coherence will not affect the population dynamics, and in particular, will have no effect on the pulsing behavior. The main detrimental process, therefore, is the loss of particles within the ladder of modes. For an implementation in superconducting circuits, in which individual modes are realized using high-$Q$ microwave cavities, such losses can be as low as $\kappa_0=10 \text{kHz}$~\cite{Frunzio2005,Reagor2013,Somoroff2023}. Given the parameters we have for the implementation, we can thus expect $\kappa_0\sim 0.1-1 \Gamma$.
We performed mean-field simulations taking such processes into account; in Fig.\ref{fig:PD_withchainloss}, we display the phase diagram we obtain, using the same parameters as in Fig.\ref{fig:phasediag} in the main text, and $\kappa_0/\Gamma$ ranging from $0.1$ to $10$. We see that the phase diagram is almost unaffected for all but the largest values of $\kappa_0$, and that, even in this case, we can clearly distinguish the three phases we discussed. Based on this mean-field analysis, we expect that none of our findings will be critically affected by the typical levels of losses one can expect in superconducting circuits. 

\section{Details on pulsing dynamics and scaling of the period}
\label{sec:appendix_period}

\begin{figure}
    \centering    \includegraphics[width=0.95\linewidth]{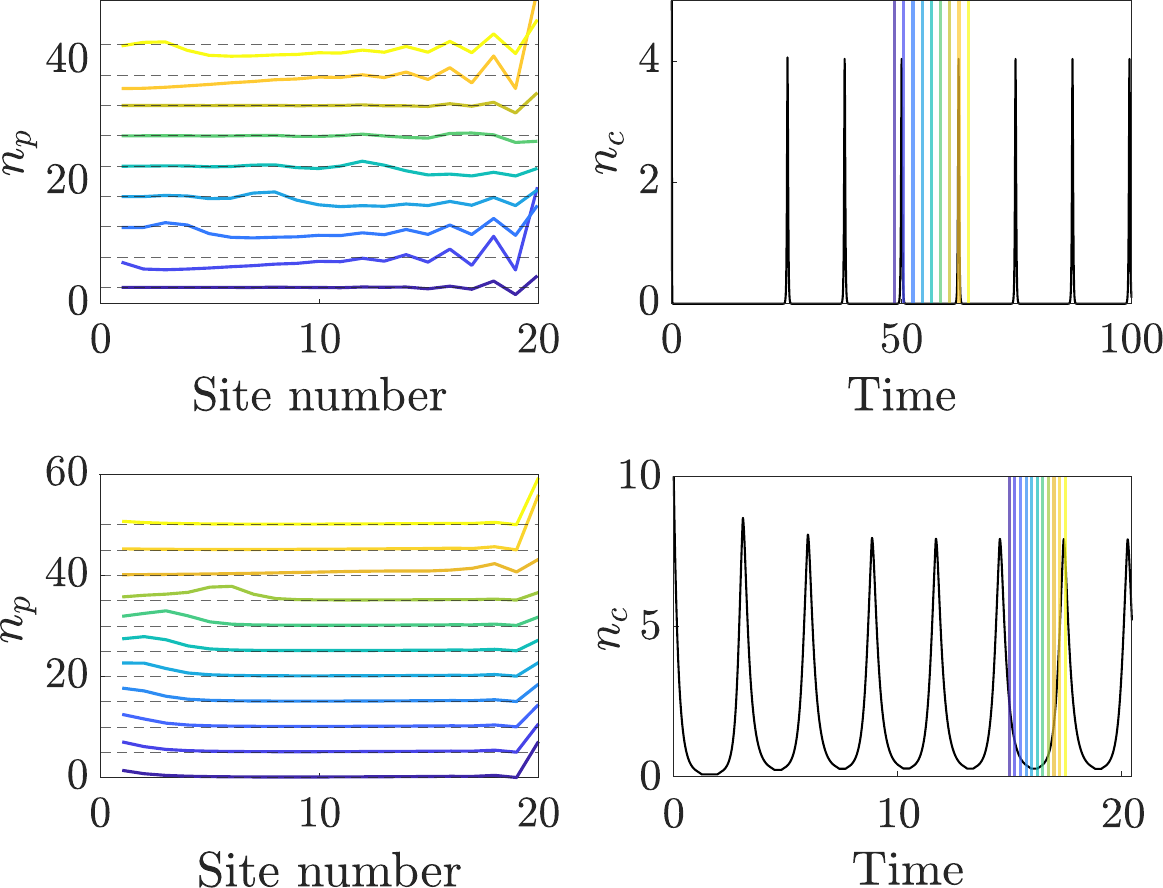}
    \caption{Left: snapshots of the cavity population distribution at different times. For better visualization, each curve is shifted with respect to the lowermost one; the dotted lines indicate the shifted "baseline" of each curve. Right: cavity population evolution, with the times at which we take the snapshot indicated by colored vertical lines. The top plot corresponds to a cavity decay rate $\kappa_c/\Gamma=150$, the bottom one to $\kappa_c/\Gamma=20$. Other parameters are $N=20$, $\kappa_\ell/\Gamma=2$, and $\chaingain/\Gamma=10$. The dynamics at high and low cavity decay is characterized by depletion and population waves, respectively.}
    \label{fig:pulsingshape}
\end{figure}

In this section, we elaborate on the dynamics of the ladder in the pulsing phase, and on the determination of the oscillation period. In Fig.\ref{fig:pulsingshape}, we show snapshots of the cavity population distribution at different times, for two different values of the cavity loss rate. These correspond to points in the pulsing phase close to the boundaries with the empty and lasing phases, respectively. In the latter case (bottom plot), we observe the kind of mechanism we described in the main text: starting from a mostly empty ladder (the rightmost site population that can be observed in the plot is a leftover from the previous pulse), we observe a packet slowly forming and propagating on the first few modes of the ladder. As described in the main text and in Appendix~\ref{sec:appendix_asip}, the speed of motion of this packet can be estimated as $c\sim \sqrt{\gamma_\textsl{g}\Gamma}$. If we assume the dynamics within the packet itself to be quasi-stationary, the current between each pair of sites is given by the steady-state condition $J_{p,p+1}=J_{ss}=\gamma_\textsl{g}$. As this packet propagates along the ladder, an increased number of modes get populated and engage in the hopping dynamics: this number can be estimated as $n_{pop}\sim c t$. The total cumulative current, then, grows like $J_{cum}\sim n_{pop} J_{ss}=\gamma_\textsl{g} c t$; this, in turn, means a larger pumping for the cavity. Once the total pumping rate exceeds the loss rate $\kappa_c$, one observes a burst-like increase in $n_c$ (right-side plot). This triggers a sudden acceleration of the packet, which quickly moves all the way through the ladder to the final site. \\

For high values of $\kappa_c$ (top plot in Fig.\ref{fig:pulsingshape}), we observe a similar, but slightly different process: instead of starting from an empty ladder, we have a non-zero filling, such that the total cavity pumping \textit{already} exceeds the threshold value $\kappa_c$. We observe, therefore, a rapid increase in the cavity population, resulting once again in an acceleration of the ladder dynamics, causing now a \textit{dip} in the ladder population on the left-hand side. To understand this, consider the dynamics of the leftmost site; in the initial configuration we considered, we have a local equilibrium between the incoming drive current $\chaingain$ and the current $\Gamma(1+\lvert\alpha_c\rvert^2) n_1(1+n_2)$ flowing to the second site. When the cavity population spikes, the latter increase, but not the former, resulting in a population loss on the first site (modes in the center of the ladder, by contrast, are initially unaffected, since \textit{both} their incoming and outgoing currents are amplified). This initial quench on the first site then quickly affects the neighboring modes, creating the depleted region we observe on the first few modes of the ladder. The decrease in population leads to a lower overall current, making the cavity population decay once more. We are therefore left with a depleted region in the ladder, which is going to propagate at the same effective speed $c$ we discussed above, until it reaches the right edge. Once it starts subsiding, the population moves back towards its initial constant filling configuration, and the current starts increasing again, restarting the cycle.\\

To summarize, these two opposite regimes of high and low $\kappa_c$ exhibit behaviors that are more or less mirror images of one another: in the former, we have a \textit{slow} creation of a \textit{population wave}, followed by \textit{rapid} propagation through the ladder; in the latter, we have a \textit{rapid} creation of a \textit{depletion wave}, followed by \textit{slow} propagation through the ladder. The slow propagation of \textit{both} kind of waves, however, occurs with the same typical speed $c\sim\sqrt{\Gamma\chaingain}$, which will dominate the overall duration of the process.

\begin{figure}[H]
    \centering    \includegraphics[width=0.95\linewidth]{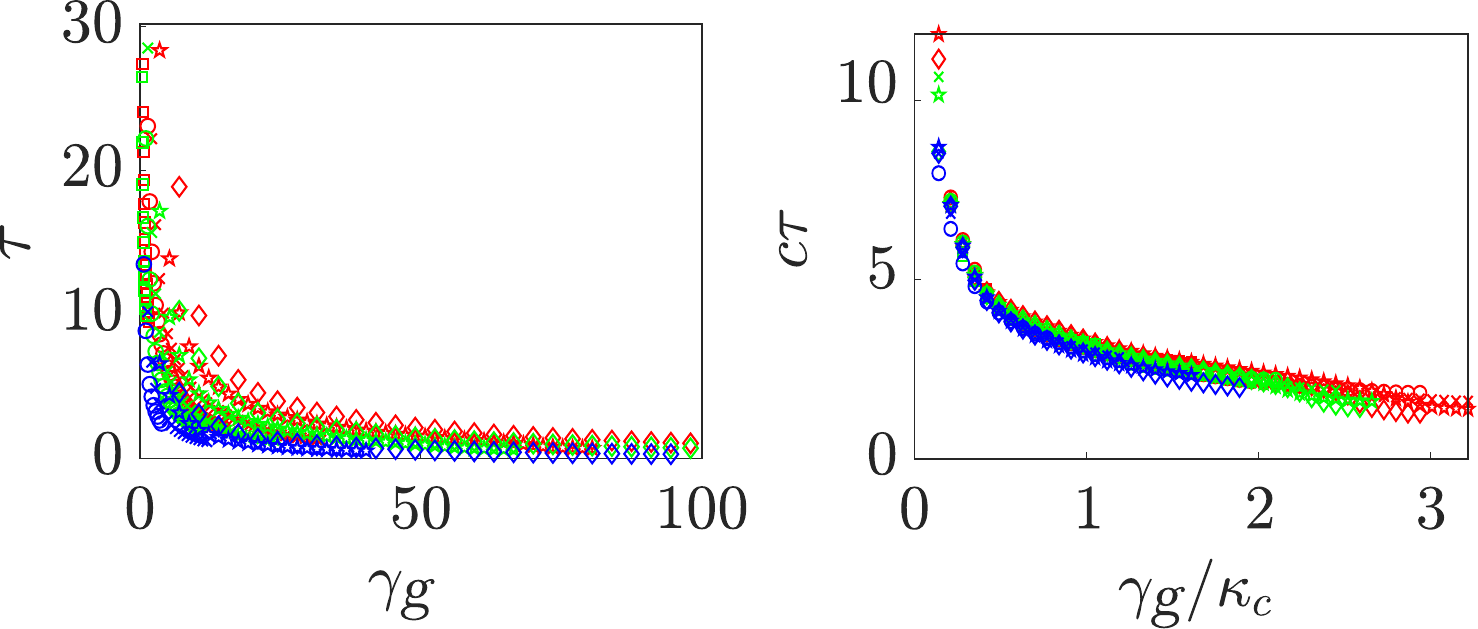}
    \caption{Left: unrescaled period versus gain rate for $N=10$ \change{(all frequencies and times are defined with respect to $\kappa_\ell$, set here to $1$)}, taking multiple values for the hopping rate and cavity loss: $\Gamma/\kappa_\ell=0.5,0.1,0.05$ (blue, green, and red symbols, respectively), and $\kappa_c/\kappa_\ell=1,5,10,25,50$ (squares, circles, crosses, stars, and diamonds, respectively). Right: rescaled period versus rescaled pumping, with the same parameters. We used this plot for the inset of Fig.\ref{fig:packet_pulsing} in the main text.}
    \label{fig:resc_full}
\end{figure}

The argument given in the main text suggests that the period, rescaled by the effective speed $c\sim \sqrt{\gamma_\textsl{g}\Gamma}$, should be compared against the rescaled gain $\gamma_\textsl{g}/\kappa_c$.
In Fig.\ref{fig:resc_full}, we display the unrescaled and rescaled periods for multiple values of the hopping rate, cavity loss rates, and gain. We observe a very good collapse of all curves, indicating that the ansatz $c\tau=f(\kappa_c/\chaingain)$ works quite well for both high and low $\kappa_c$. However, a more detailed analysis of the function $f$ reveals that it deviates from the behavior $f(\kappa_c/\chaingain)=\kappa_c/\chaingain$ that one would expect from the argument given in the main text. The precise behavior of this function is not fully clear and will be investigated in future works.

\section{Monte-Carlo simulations} \label{sec:appendix_numerics}

In the absence of any additional Hamiltonian terms, the master equation in Eq.~\eqref{eq:Lindblad} is diagonal in the Fock basis $|\{ \vec n\}\rangle=|n_c;n_1,n_2..,n_N\rangle$,where $n_c$ and $n_{1,..N}$ denote the population of the cavity and the modes in the ladder (i.e., the gain medium), respectively. Therefore, the diagonal elements of the density operator, $P(\{\vec n\},t)= \langle \{ \vec n\}| \hat\rho(t)|\{ \vec n\}\rangle$, which describe the probabilities of different particle configurations, evolve independently from the off-diagonal elements, according to the equation:

\begin{widetext}
 \begin{align*}
  \dot P= &  \Gamma \sum_p n_c n_{p+1} (1+n_{p}) P(\{\vec n+\vec\mu_p\}) -  (1+n_c)n_{p} (1+n_{p+1}) P\\
  +& \sum_{\lambda=\{c,1,N\}} \kappa_\lambda \Big\{(1+n_\lambda)P(\{\vec n+\vec\epsilon_{\lambda}\})-n_\lambda P\Big\}
  +\gamma_\textsl{g} \Big\{n_1 P(\{\vec n-\vec\epsilon_{1}\})-(1+n_1)P \Big\}
   \end{align*} 
\end{widetext}
Here, $\epsilon_p^j=\delta_{pj}$, and $\vec\mu=\vec\epsilon_p-\vec\epsilon_{p+1}-\vec\epsilon_c$; $\{\vec n+\vec\epsilon_p-\vec\epsilon_{p+1}-\vec\epsilon_c\}$ is the configuration obtained from $\{\vec n\}$ by removing one excitation in the cavity \textit{and} on site $p+1$, and adding one on site $p$. We also used a short notation $P=P(\{\vec n\})$, and omitted time dependence to lighten up the equation.\\

To study this dynamics, we sampled the probability according to a Monte-Carlo algorithm. The boson numbers $\{n_c;n_{1...N}(t)\}$ are treated as stochastic variables, which during an infinitesimal time step evolve according to 
 \begin{align} \nonumber
 d n_{q=1..N}&=dH_{q-1}-dH_q+\delta_{q1}(dG_1-dL_1)-\delta_{qN}dL_N\\ 
 d n_c&=\sum_{q=1}^N dH_q-dL_c
 \label{eq:stochmonte}
 \end{align}
 Where the $dH_q$, $dG_\lambda$, and $dL_\lambda$ are independent random variables, taking binary values $\{0,1\}$, and indicating that a hopping, gain, or loss event, respectively, took place. In a short enough time interval $dt$, the respective probabilities that these variables assume value $1$ are $p(dH_q=1)=\Gamma (1+n_c)n_{q+1}(1+n_q)dt$, $p(dG_1=1)=\gamma_\textsl{g}(1+n_1) dt$, and $p(dL_\lambda=1)=\kappa_\lambda n_\lambda dt$.

The simulation is performed using a standard Gillespie algorithm \cite{Gillespie1977}.
By starting from a given initial configuration, $\{n_p(t=0)\}$, and evolving a total number of $\mathcal{N}_t$ stochastic trajectories in time, we can approximate the expectation value of any function of operators $\hat n_p$ by an ensemble average. For example,
 \begin{equation}
 \langle \hat n_p \hat n_q\rangle(t) \simeq \frac{1}{\mathcal{N}_t} \sum_{i=1}^{\mathcal{N}_t} n_p(t) n_q(t)=:\langle n_p(t) n_q(t)\rangle. 
 \end{equation} 
 This method becomes exact in the limit $\mathcal{N}_t\rightarrow \infty$, and accounts for both average populations and population fluctuations.

\section{Coherence resonance with other pumping mechanisms}
\label{sec:pumpmech_appendix}

\begin{figure}[H]
\centering \includegraphics[width=0.95\linewidth]{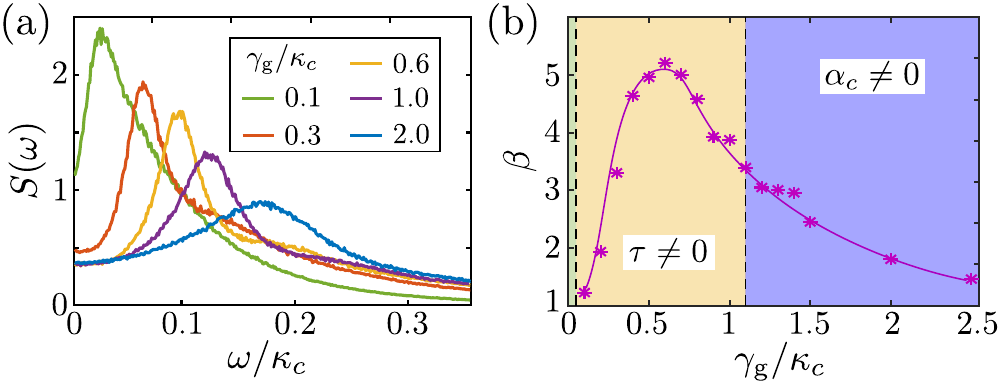}
\caption{Cavity population (left), noise spectra (top right), and coherence factor (bottom right), for pure gain input ($\zeta=1$). All other parameters are the same as in Fig.\ref{fig:SCR} of the main text.}
    \label{fig:coherence_puregain}
\end{figure}

In the main paper, we have mostly focused on a pumping mechanism with $\zeta=1/2$, which corresponds to equal gain and loss on the first ladder mode. 
Similar effects, however, can also be achieved with other pumping mechanisms. In Fig.\ref{fig:coherence_puregain}, we show the trajectories, spectra, and coherence factor obtained with $\zeta=1$, which corresponds to pure gain in the first ladder site. 
The same basic features (presence of bursts in the cavity population, whose coherence peaks for an intermediate value of the gain) can again be observed.

\section{Definition of stochastic and coherence resonances in the quantum regime}
\label{sec:appendix_discussion}

\change{Stochastic resonance is typically studied for signals with a \textit{continuous} amplitude, for which a pure stochastic drive takes on the form of a signal oscillating around $0$ without time correlation. The present system, however, is best described in terms of chain populations, that adopt \textit{discrete}, \textit{positive} values. For such signals, a random input can be a source injecting or extracting particles at random times. Such processes, however, \textit{must} be biased towards injection, in order to preserve the positivity of the signal. To illustrate this point, let us consider the infinite-temperature gain process described by \eqref{eq:injection_lindbladian} with $\zeta=1/2$. This process can be equivalently described by a Langevin equation for the creation operator on the first site: 
\begin{equation}
\dot{a}_1=\sqrt{\frac{\gamma_\textsl{g}}{2}} \left(a_{in,1}(t)+a^\dag_{in,2}(t)\right),
\end{equation}
where $a_{in,x}(t)$ are input fields describing the initial state of the environment, which satisfy $[a_{in,x}(t), a^\dag_{in,y}(t')]=\delta_{xy}\delta(t-t')$, and $\langle a_{in,x}(t)\rangle=0$. Thus, from the perspective of the field \textit{amplitude} $\langle a_1\rangle$, the mechanism we considered corresponds to a \textit{purely} stochastic driving, with \textit{no} average gain. By contrast, at the level of the population $n_1=\langle a^\dag_1 a_1\rangle$, this mechanism is realized by a Monte-Carlo process whereby a particle is injected on the first site on each time-step with a probability $p_i\propto \gamma_\textsl{g} (1+n_1)$, and removed with a probability $p_r\propto \gamma_\textsl{g} n_1$, with $n_1$ the number of particles already present on the first site. Comparing these two rates, it is evident that the particle number will indeed experience stochastic fluctuations, but with the aforementioned bias towards injection $p_i-p_r\propto \gamma_\textsl{g}$.}\\

\change{We can then ask the following question: what is the mean-field equivalent of such a process, when the noise is ``taken away"? If we look at the evolution of the particle number, this is just a driving at a constant rate $dn=\gamma_\textsl{g} dt$. By contrast, if we focus on the \textit{amplitude}, the driving is purely stochastic; the mean-field dynamics would then be obtained by setting $\gamma_\textsl{g}=0$, that is, not having a drive at all. The point here is that there is a fundamental ambiguity in identifying stochastic and deterministic contributions when dealing with quantum signals: depending on the definition we adopt, the process \eqref{eq:injection_lindbladian} may be seen as a mean-field contribution plus noise, or as pure noise. In Fig.~\ref{fig:SCR}, we have adopted the first convention and compared the MC results to mean-field equations \eqref{eq:MF_casc} with a drive rate $\gamma^{MF}_{\textsl{g}}=\gamma_\textsl{g}$. We then concluded that we already had oscillations at the mean-field level, which are modified and, in some regime, amplified by the noise, which is more reminiscent of SR than CR. If one takes the second perspective, by contrast, the stochastic process corresponds to \textit{pure} noise, and the corresponding mean-field regime is $\gamma^{MF}_{\textsl{g}}=0$, which never leads to oscillations. In that case, one could then conclude that the behavior observed in Fig.\ref{fig:SCR} is generated entirely by noise, and thus constitutes a true instance of CR. This is in part semantics, and we will not push further this question here. The point of this discussion, however, was to emphasize that the notions of a noisy and deterministic signal, and the resulting notions of stochastic and coherence resonance, can become somewhat ill-defined when dealing with quantum operators, and that care should be taken in applying these notions.}

\section{Implementation: details}\label{app:Implementation}

\begin{figure}
    \centering       
     \includegraphics[width=\columnwidth]{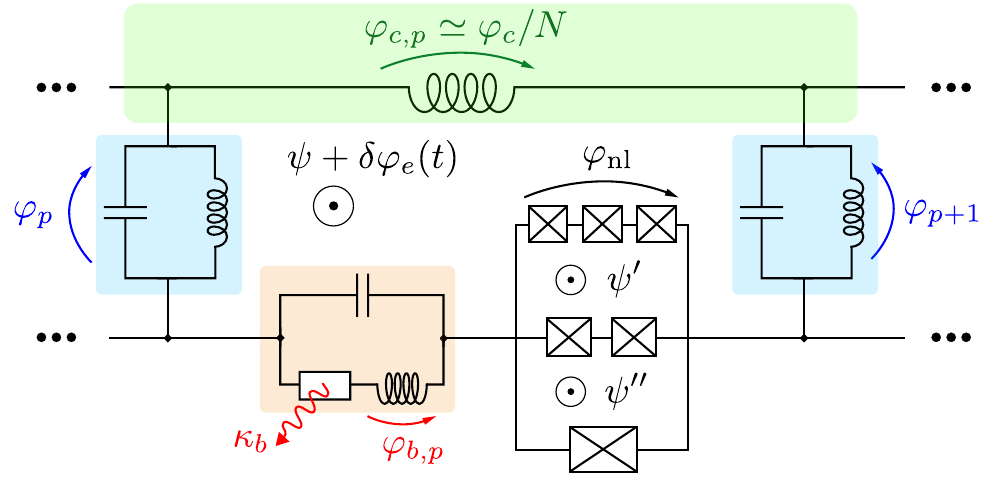}
    \caption{Detailed circuit layout. The plot shows a zoom of one of the unit cells of the full circuit in Fig.~\ref{fig:circuit}(a), where two neighboring ladder modes are coupled to the cavity mode and to an additional dissipative waste mode with phase variable $\varphi_{b,p}$ via a SNAIL-type coupler. The latter is composed of Josephson junctions arranged on three parallel branches threaded by the normalized magnetic fluxes $\psi'$ and $\psi''$. The full circuit in Fig.~\ref{fig:circuit}(a) is obtained by repeating this basic cell in series and adding an additional capacitor $C_c$ for the cavity mode. See text for more details. 
    }
     \label{fig:circuit_detailed}
 \end{figure}

In Fig.~\ref{fig:circuit_detailed} , we show a  detailed layout of the proposed circuit for implementing the master equation given in Eq.~\eqref{eq:Lindblad}. Specifically, this plot focuses on a single unit cell of the full circuit depicted in Fig.~\ref{fig:circuit}(a), which realizes the dissipative three-mode process with dissipator $\mathcal{D}[\aop_p \aop_{p+1}^\dag \cop^\dag]$. 
Within this unit cell, the two neighboring ladder modes with flux variables $\phi_p$ and $\phi_{p+1}$, an auxiliary waste mode with flux variable $\phi_{b,p}$, and the cavity mode with a local flux $\phi_{c,p}$ are coupled via a nonlinear, SNAIL-type~\cite{Frattini2017} element. To achieve optimal tunability, this nonlinear coupler has three branches containing one, two, and three Josephson junctions, with Josephson energies $E_J$, $\alpha_2 E_J$, and $\alpha_3 E_J$, respectively. In the following, we denote by $\varphi_x=\phi_x/\phi_0$ the dimensionless phase variables associated with all fluxes, where $\phi_0$ is the reduced flux quantum, $\phi_0=\hbar/(2e)$. Note that here we assume that the lasing cavity is realized by a lumped-element resonator with a fundamental frequency $\omega_c$ that is well-separated from higher excited modes. This means that local variations of the flux across the whole ladder can be neglected. Therefore, for each unit cell, we can set $\varphi_{c,p}\simeq \varphi_c/N$, where $\varphi_c$ is the phase variable of the cavity mode. Similarly, we ignore intrinsic, high-frequency excitations of the SNAIL coupler, such that its energy can be described by a single degree of freedom,  $\varphi_{\rm nl}$.

Each unit cell forms a closed loop, and we denote by $\psi+\delta\varphi_e(t)$, $\psi'$ and $\psi''$ the dimensionless external fluxes, which are threading the whole loop and the branches of the SNAIL elements, respectively [(see Fig.~\ref{fig:circuit_detailed}). The flux quantization condition then enforces the constraint $\varphi_{\text{nl}}=\psi+\varphi_{{\rm tot},p}$, where
\begin{equation}
\varphi_{{\rm tot},p}=\delta\varphi_{e}(t)+\varphi_{p}-\varphi_{p+1}+\varphi_{c,p}-\varphi_{b,p}.
\end{equation}
Taking this condition into account, we can follow the usual quantization procedure to obtain the Hamiltonian of the whole circuit, which reads
\begin{align}\label{eq:CircuitHamiltonian}
H&= \omega_c \cop^\dag \cop + \sum_{p=1}^N \omega_p \aop^\dag_p \aop_p + \sum_{p=1}^{N-1} \left(\omega_{b}  b_p^\dag  b_p +  H_{\rm nl}^{(p)}\right).
\end{align}
Here, 
\begin{align}
H_{\rm nl}^{(p)} =\, & -E_J\left[ 3\alpha_3\cos\left(\frac{\psi+\varphi_{{\rm tot},p}}{3}\right) \right.  \nonumber \\  &\left.+2\alpha_2\cos\left(\frac{\chi+\varphi_{{\rm tot},p}}{2}\right)+\cos\left(\theta+\varphi_{{\rm tot},p}\right)\right],
\end{align}
is the Hamiltonian of the $p$-th nonlinear coupler with $\chi=\psi'+\psi$ and $\theta=\psi''+\psi'+\psi$. In Eq.~\eqref{eq:CircuitHamiltonian} we have introduced bosonic annihilation and creation operators according to  
\begin{equation}
\varphi_p=\sqrt{\frac{Z_p}{Z_0}}(\aop_p+\adag_p),
\end{equation}
where $Z_p=\sqrt{L_p/C_p}$ is the impedance of the ladder mode resonators and $Z_0=\hbar/e^2=4.1 \text{k}\Omega$. Analogous expressions apply to the cavity mode and the waste modes, with their respective impedances $Z_c$ and $Z_b$.

To proceed, we expand the coupling as
\begin{align}
H_{\rm nl}^{(p)} = -\sum_{n=0}^\infty \frac{B_{n}}{n!}\varphi^{n}_{{\rm tot},p},
\end{align}
where
\begin{align*}
    &B_{2p+1}=(-1)^p\left[\frac{\alpha_3}{3^{2p}} \sin(\psi/3)+\frac{\alpha_2}{2^{2p}}\sin(\chi/2)+ \sin(\theta)\right],\\
    &B_{2p+2}=(-1)^{p}\left[\frac{\alpha_3}{3^{2p}} \cos(\psi/3)+\frac{\alpha_2}{2^{2p}}\cos(\chi/2)+ \cos(\theta)\right].
\end{align*}
We are interested in the fifth-order term, and to make this contribution dominant, we need to tune the fluxes and energies of the junctions appropriately to cancel the lower-order contributions. In particular, it is possible to cancel out the first, second, and third order, by imposing the conditions
\begin{align*}    \alpha_3\sin(\psi/3)+\alpha_2\sin(\chi/2)+\sin(\theta)=0,\\ \frac{\alpha_3}{3}\cos(\psi/3)+\frac{\alpha_2}{2}\cos(\chi/2)+\cos(\theta)=0,\\
    \frac{\alpha_3}{9}\sin(\psi/3)+\frac{\alpha_2}{4}\sin(\chi/2)+\sin(\theta)=0.
\end{align*}
These can be met by an appropriate choice of fluxes and energies, and we will present examples of such choices below. 
Once these conditions are met, the lowest non-vanishing contributions in the expansion of $H_{\rm nl}^{(p)}$ are
\begin{align}\label{eq:Hnl_expanded}
H_{\rm nl}^{(p)} \simeq H_{\rm Kerr}^{(p)} -\frac{2}{3}\sin(\theta)E_J\frac{\varphi^{5}_{{\rm tot},p}}{5!},
\end{align}
where $H_{\rm Kerr}^{(p)}$ is the residual fourth-order contribution, which can lead to Kerr-type frequency shifts that we address below. The second term in Eq.~\eqref{eq:Hnl_expanded} is the fifth-order term of interest.

By expanding $\varphi^{5}_{{\rm tot},p}$ in terms of annihilation and creation operators for all the modes involved, we obtain various multi-photon processes. In general, these are, however, non-resonant and energetically suppressed. To resonantly enhance a specific process, we focus on the contribution 
$\delta\varphi_e(t) \varphi_{b,p}\varphi_{c,p}\varphi_{p+1}\varphi_p$ and assume a periodically modulated external flux $\delta\varphi_e(t)=\delta\varphi_e\cos(\omega_e t)$, where the modulation frequency satisfies $\omega_e=\omega_{p+1}-\omega_p+\omega_{c}+\omega_{b}$. This choice makes processes of the type $\aop_p \adag_{p+1} \cdag b_p^\dag $ resonant, and we obtain a dominant interaction of the form 
\begin{align}\label{eq:Hnl_ideal}
H_{\rm nl}^{(p)}\approx V_p=  g \left( \aop_p \adag_{p+1} \cdag  b_p^\dag + \aop^\dag_p \aop_{p+1} \cop b_p \right).
\end{align}
This process will transfer a photon from mode $p$ to $p+1$, while creating excitations in the cavity and waste modes. By assuming, for simplicity, an equal impedance $Z_x\simeq Z$ for all modes, the corresponding coupling strength is given by 
\begin{align}
    g=\frac{2E_J\sin(\theta)}{3\hbar N}\delta\varphi_e\left(\frac{Z}{Z_0}\right)^2.
\end{align}
In a final step, we can follow the usual procedure and adiabatically eliminate the dissipative waste mode to derive an effective master equation for the remaining degrees of freedom~\cite{GardinerZoller2004}. This derivation, which is valid when $\kappa_b\gg g$, results in a dissipator $\Gamma \mathcal{D}[C]$ with jump operator $C=\aop_p \adag_{p+1} \cdag$ and a rate $\Gamma=4g^2/\kappa_b$, as assumed in our model in Eq.~\eqref{eq:Lindblad}.

Let us now return to the remaining fourth-order processes $\varphi^4_{\rm tot,p}$, contained in $H_{\rm Kerr}^{(p)}$. It is not possible to tune the fluxes and Josephson energies to remove these contributions while suppressing all lower-order contributions as well. Further, some of these fourth-order terms correspond to Kerr and cross-Kerr interactions, which conserve the photon number in each mode and lead to static energy shifts that cannot be eliminated by a rotating-wave approximation. These terms are of the form 
\begin{align} \label{eq:kerr_proc}
H_{\rm Kerr}^{(p)}&=B_4E_J\left(\frac{Z}{Z_0}\right)^2 \sum_{x,y}\alpha_{x,y} n'_x  n'_y,
\end{align}
where the indices $x,y\in\{p,p+1,c,(b,p)\}$ run over the four involved modes 
and $\alpha_{x,y}=1/4$ for $x=y$ and $1/2$ otherwise. Further, we have set  $ n'_c=\adag_c\aop_c/N^2$ and $ n'_x=\adag_x\aop_x$ for all other modes. 

The processes in Eq.~\eqref{eq:kerr_proc} are present for all choices of frequencies in our setup. While they conserve the boson numbers and therefore have \textit{no effect} on the dissipative dynamics in the final model in Eq.~\eqref{eq:Lindblad}, they impact the resonance condition assumed in Eq.~\eqref{eq:Hnl_ideal}. Therefore, these Kerr-shifts should be small compared to $\kappa_b$, which determines the width of the resonance. Additional unwanted processes may also arise as a result of accidental resonances. For example, if $\omega^3_p\simeq \omega_{b}$, the processes $\aop^3_{p} b_p^\dag$ becomes resonant and results in an additional three-photon loss process. Therefore, such accidental resonances must be avoided. For the parameters given in Table \ref{tab:my_label} below, we have explicitly verified these conditions for up to $N=5$ and found that all unwanted processes up to the fifth-order expansion are out of resonance by at least $\delta_{\rm min} /(2\pi)\approx 200$ MHz. Altogether, we find that both the Kerr interactions as well as unwanted resonances can be neglected when the hierarchy 
\begin{equation}\label{eq:parameter_hierarchy}
 \frac{1}{2}B_4E_J\left(\frac{Z}{Z_0}\right)^2 \times {\rm max}\{\bar{n},\bar n_c/N^2\}\lesssim \kappa_b\ll \delta_{\rm min},
 \end{equation}
is satisfied. Here, $\bar{n}$ and $\bar{n}_c$ are the typical photon numbers of the ladder modes and the cavity mode, respectively.  

\begin{table}[t]
    \centering
    \begin{tabular}{|c|c||c|c|}
    \hline
        $\omega_1/(2\pi) $ & $4.7$ \text{GHz} & $\kappa_b/(2\pi)$ & $30$ \text{MHz} \\
        \hline
        $\Delta\omega/(2\pi)$ & $300$ \text{MHz} & $\kappa_c/(2\pi)$ & $0.02-1$ \text{MHz} \\
        \hline
        $\omega_c/(2\pi)$  & $3.6$ \text{GHz} &  $Z$ & $160\, \Omega$\\
        \hline
        $\omega_{b}/(2\pi)$ & $10.7$ \text{GHz} &  $\delta \varphi_e$ & $0.25$\\
        \hline
        $E_J/h$ & $50$ \text{GHz} & $\kappa_0/(2\pi)$ & $20$ \text{kHz} \\
        \hline
    \end{tabular}
    \caption{Parameter example for the realization of a bosonic avalanche laser, using the circuit layout shown in Fig.~\ref{fig:circuit}(a) and Fig.~\ref{fig:circuit_detailed} for $N=5$ ladder modes. The frequencies of the ladder modes are chosen as $\omega_p= \omega_1-(p-1)\Delta \omega$ and for all modes the same impedance $Z$ is assumed.}
    \label{tab:my_label}
\end{table}

As a specific example, we set $\alpha_2=2.4$, $\alpha_3=2.1$, $\sin(\chi/2)=0.88$, $\sin(\psi/3)=0.85$ and $\sin(\theta)=0.33$ to cancel all low-order contributions, as discussed above.  
For this choice, $N=5$ and the other parameters listed in Table~\ref{tab:my_label}, we obtain $B_4\simeq 0.75$,  $B_4E_J\left(Z/Z_0\right)^2/2\simeq 2\pi\times 30$ MHz and $g/(2\pi)\simeq 850$ kHz. For a decay rate $\kappa_b/(2\pi)=30$ MHz, we then obtain a hopping rate of $\Gamma/(2\pi)\simeq 100$ kHz. This rate exceeds the bare losses of a high-Q superconducting resonator mode, $\kappa_0/(2\pi)\approx 10-100$ kHz~\cite{Frunzio2005,Reagor2013,Somoroff2023}, while at the same time the condition in Eq.~\eqref{eq:parameter_hierarchy} is satisfied for low photon numbers. Note that depending on the regime of operation, the impedance of the cavity $Z_c$ mode could be further increased to enhance the coupling $g\sim \sqrt{Z_c}$, without affecting the most detrimental Kerr interactions between the ladder modes. This and other parameter optimizations can be used to achieve similar conditions also for $N\gtrsim 10$.

\bibliography{Ref_AvalancheLaser.bib}

@article{garbe_bosonic_2024,
title = {The bosonic skin effect: Boundary condensation in asymmetric transport},
author = {Garbe, L. and Minoguchi, Y. and Huber, J. and Rabl, P.},
journal = {SciPost Phys.},
volume = {16},
number = {1},
year = {2024},
pages = {029},
url = {https://scipost.org/SciPostPhys.16.1.029},
doi = {10.21468/SciPostPhys.16.1.029},
}

@article{Minoguchi2025,
  title = {Unified Interface Model for Dissipative Transport of Bosons and Fermions},
  author = {Minoguchi, Y. and Huber, J. and Garbe, L. and Gambassi, A. and Rabl, P.},
  journal = {Phys. Rev. Lett.},
  volume = {134},
  issue = {20},
  pages = {207102},
  numpages = {8},
  year = {2025},
  month = {May},
  publisher = {American Physical Society},
  doi = {10.1103/PhysRevLett.134.207102},
  url = {https://link.aps.org/doi/10.1103/PhysRevLett.134.207102}
}

@article{lindner_effects_2004,
	title = {Effects of noise in excitable systems},
	volume = {392},
	issn = {0370-1573},
	url = {https://www.sciencedirect.com/science/article/pii/S0370157303004228},
	doi = {10.1016/j.physrep.2003.10.015},
	number = {6},
	urldate = {2024-08-10},
	journal = {Phys. Rep.},
	author = {Lindner, B. and Garcia-Ojalvo, J. and Neiman, A. and Schimansky-Geier, L.},
	month = mar,
	year = {2004},
	pages = {321--424},
}

@article{pikovsky_coherence_1997,
	title = {Coherence {Resonance} in a {Noise}-{Driven} {Excitable} {System}},
	volume = {78},
	url = {https://link.aps.org/doi/10.1103/PhysRevLett.78.775},
	doi = {10.1103/PhysRevLett.78.775},
	number = {5},
	urldate = {2024-08-14},
	journal = {Phys. Rev. Lett.},
	author = {Pikovsky, Arkady S. and Kurths, Jürgen},
	month = feb,
	year = {1997},
	pages = {775--778},
}

@article{kato_quantum_2021,
	title = {Quantum coherence resonance},
	volume = {23},
	issn = {1367-2630},
	url = {https://dx.doi.org/10.1088/1367-2630/abf1d7},
	doi = {10.1088/1367-2630/abf1d7},
	number = {4},
	urldate = {2024-08-14},
	journal = {New J. Phys.},
	author = {Kato, Yuzuru and Nakao, Hiroya},
	month = apr,
	year = {2021},
	pages = {043018},
}

@article{lofstedt_quantum_1994,
	title = {Quantum stochastic resonance},
	volume = {72},
	url = {https://link.aps.org/doi/10.1103/PhysRevLett.72.1947},
	doi = {10.1103/PhysRevLett.72.1947},
	number = {13},
	urldate = {2024-08-14},
	journal = {Phys. Rev. Lett.},
	author = {Löfstedt, R. and Coppersmith, S. N.},
	month = mar,
	year = {1994},
	pages = {1947--1950},
}

@article{grifoni_coherent_1996,
	title = {Coherent and {Incoherent} {Quantum} {Stochastic} {Resonance}},
	volume = {76},
	url = {https://link.aps.org/doi/10.1103/PhysRevLett.76.1611},
	doi = {10.1103/PhysRevLett.76.1611},
	number = {10},
	urldate = {2024-08-14},
	journal = {Phys. Rev. Lett.},
	author = {Grifoni, Milena and Hänggi, Peter},
	month = mar,
	year = {1996},
	pages = {1611--1614},
}

@article{grifoni_quantum_1996,
	title = {Quantum tunneling and stochastic resonance},
	volume = {53},
	url = {https://link.aps.org/doi/10.1103/PhysRevE.53.5890},
	doi = {10.1103/PhysRevE.53.5890},
	number = {6},
	urldate = {2024-08-14},
	journal = {Phys. Rev. E},
	author = {Grifoni, Milena and Hartmann, Ludwig and Berchtold, Sabine and Hänggi, Peter},
	month = jun,
	year = {1996},
	pages = {5890--5898},
}

@article{mompo_coherence_2018,
	title = {Coherence {Resonance} and {Stochastic} {Resonance} in an {Excitable} {Semiconductor} {Superlattice}},
	volume = {121},
	url = {https://link.aps.org/doi/10.1103/PhysRevLett.121.086805},
	doi = {10.1103/PhysRevLett.121.086805},
	number = {8},
	urldate = {2024-08-14},
	journal = {Phys. Rev. Lett.},
	author = {Mompo, Emanuel and Ruiz-Garcia, Miguel and Carretero, Manuel and Grahn, Holger T. and Zhang, Yaohui and Bonilla, Luis L.},
	month = aug,
	year = {2018},
	pages = {086805},
}

@article{gang_stochastic_1993,
	title = {Stochastic resonance without external periodic force},
	volume = {71},
	url = {https://link.aps.org/doi/10.1103/PhysRevLett.71.807},
	doi = {10.1103/PhysRevLett.71.807},
	number = {6},
	urldate = {2024-08-15},
	journal = {Phys. Rev. Lett.},
	author = {Gang, Hu and Ditzinger, T. and Ning, C. Z. and Haken, H.},
	month = aug,
	year = {1993},
	pages = {807--810},
}

@article{liew_proposal_2013,
	title = {Proposal for a {Bosonic} {Cascade} {Laser}},
	volume = {110},
	url = {https://link.aps.org/doi/10.1103/PhysRevLett.110.047402},
	doi = {10.1103/PhysRevLett.110.047402},
	number = {4},
	urldate = {2024-12-20},
	journal = {Phys. Rev. Lett.},
	author = {Liew, T. C. H. and Glazov, M. M. and Kavokin, K. V. and Shelykh, I. A. and Kaliteevski, M. A. and Kavokin, A. V.},
	year = {2013},
	pages = {047402},
}

@article{bonilla_nonlinear_2024,
	title = {Nonlinear {Charge} {Transport} and {Excitable} {Phenomena} in {Semiconductor} {Superlattices}},
	volume = {26},
	copyright = {http://creativecommons.org/licenses/by/3.0/},
	issn = {1099-4300},
	url = {https://www.mdpi.com/1099-4300/26/8/672},
	doi = {10.3390/e26080672},
	number = {8},
	urldate = {2025-06-26},
	journal = {Entropy},
	author = {Bonilla, Luis L. and Carretero, Manuel and Mompó, Emanuel},
	month = aug,
	year = {2024},
	pages = {672},
}

@article{wellens_stochastic_2004,
	title = {Stochastic resonance},
	volume = {67},
	doi = {10.1088/0034-4885/67/1/R02},
	journal = {Rep. Prog. Phys.},
	author = {Wellens, Thomas and Shatokhin, V. and Buchleitner, Andreas},
	month = jan,
	year = {2004},
	pages = {45},
}

@article{kavokin_bosonic_2016,
	title = {Bosonic lasers: {The} state of the art ({Review} {Article})},
	volume = {42},
	issn = {1063-777X},
	shorttitle = {Bosonic lasers},
	url = {https://doi.org/10.1063/1.4948614},
	doi = {10.1063/1.4948614},
	number = {5},
	urldate = {2025-07-18},
	journal = {Low Temp. Phys.},
	author = {Kavokin, Alexey and Liew, Timothy C. H. and Schneider, Christian and Höfling, Sven},
	month = may,
	year = {2016},
	pages = {323--329},
}

@article{liew_quantum_2016,
	title = {Quantum statistics of bosonic cascades},
	volume = {18},
	issn = {1367-2630},
	url = {https://dx.doi.org/10.1088/1367-2630/18/2/023041},
	doi = {10.1088/1367-2630/18/2/023041},
	number = {2},
	urldate = {2025-07-18},
	journal = {New J. Phys.},
	author = {Liew, T C H and Rubo, Y G and Sheremet, A S and Liberato, S De and Shelykh, I A and Laussy, F P and Kavokin, A V},
	month = feb,
	year = {2016},
	pages = {023041},
}

@article{kaliteevskii_double-boson_2013,
	title = {Double-boson stimulated terahertz emission in a polariton cascade laser},
	volume = {39},
	issn = {1090-6533},
	url = {https://doi.org/10.1134/S1063785013010148},
	doi = {10.1134/S1063785013010148},
	number = {1},
	urldate = {2025-07-18},
	journal = {Tech. Phys. Lett.},
	author = {Kaliteevskii, M. A. and Ivanov, K. A.},
	month = jan,
	year = {2013},
	pages = {91--94},
}

@article{kaliteevski_single_2014,
	title = {Single and double bosonic stimulation of {THz} emission in polaritonic systems},
	volume = {4},
	copyright = {2014 The Author(s)},
	issn = {2045-2322},
	url = {https://www.nature.com/articles/srep05444},
	doi = {10.1038/srep05444},
	number = {1},
	urldate = {2025-07-18},
	journal = {Sci. Rep.},
	author = {Kaliteevski, M. A. and Ivanov, K. A. and Pozina, G. and Gallant, A. J.},
	month = jun,
	year = {2014},
	pages = {5444},
}

@article{li_stochastic_2024,
	title = {Stochastic resonance of spinor condensates in optical cavity},
	volume = {67},
	issn = {1869-1927},
	url = {https://doi.org/10.1007/s11433-023-2278-2},
	doi = {10.1007/s11433-023-2278-2},
	number = {3},
	urldate = {2025-07-21},
	journal = {Sci. China: Phys. Mech. Astron.},
	author = {Li, Zheng-Chun and Fan, Bixuan and Zhou, Lu and Zhang, Weiping},
	month = jan,
	year = {2024},
	pages = {233011},
}

@article{huelga_stochastic_2007,
	title = {Stochastic {Resonance} {Phenomena} in {Quantum} {Many}-{Body} {Systems}},
	volume = {98},
	url = {https://link.aps.org/doi/10.1103/PhysRevLett.98.170601},
	doi = {10.1103/PhysRevLett.98.170601},
	number = {17},
	urldate = {2025-07-21},
	journal = {Phys. Rev. Lett.},
	author = {Huelga, Susana F. and Plenio, Martin B.},
	month = apr,
	year = {2007},
	pages = {170601},
}

@article{rivas_stochastic_2009,
	title = {Stochastic resonance phenomena in spin chains},
	volume = {69},
	issn = {1434-6036},
	url = {https://doi.org/10.1140/epjb/e2009-00108-5},
	doi = {10.1140/epjb/e2009-00108-5},
	number = {1},
	urldate = {2025-07-21},
	journal = {Eur. Phys. J. B},
	author = {Rivas, A. and Oxtoby, N. P. and Huelga, S. F.},
	month = may,
	year = {2009},
	pages = {51--57},
}

@article{hussein_spectral_2020,
	title = {Spectral {Properties} of {Stochastic} {Resonance} in {Quantum} {Transport}},
	volume = {125},
	url = {https://link.aps.org/doi/10.1103/PhysRevLett.125.206801},
	doi = {10.1103/PhysRevLett.125.206801},
	number = {20},
	urldate = {2025-07-21},
	journal = {Phys. Rev. Lett.},
	author = {Hussein, Robert and Kohler, Sigmund and Bayer, Johannes C. and Wagner, Timo and Haug, Rolf J.},
	month = nov,
	year = {2020},

	pages = {206801},
}

@article{xie_interference_2018,
	title = {Interference effect in optomechanical stochastic resonance},
	volume = {98},
	url = {https://link.aps.org/doi/10.1103/PhysRevE.98.052202},
	doi = {10.1103/PhysRevE.98.052202},
	number = {5},
	urldate = {2025-07-21},
	journal = {Phys. Rev. E},
	author = {Xie, Min and Fan, Bixuan and He, Xiaoli and Chen, Qingqing},
	month = nov,
	year = {2018},
	pages = {052202},
}

@article{whitney_temperature_2011,
	title = {Temperature {Can} {Enhance} {Coherent} {Oscillations} at a {Landau}-{Zener} {Transition}},
	volume = {107},
	url = {https://link.aps.org/doi/10.1103/PhysRevLett.107.210402},
	doi = {10.1103/PhysRevLett.107.210402},
	number = {21},
	urldate = {2025-07-21},
	journal = {Phys. Rev. Lett.},
	author = {Whitney, Robert S. and Clusel, Maxime and Ziman, Timothy},
	month = nov,
	year = {2011},
	pages = {210402},
}

@article{hanze_quantum_2021,
	title = {Quantum stochastic resonance of individual {Fe} atoms},
	volume = {7},
	url = {https://www.science.org/doi/full/10.1126/sciadv.abg2616},
	doi = {10.1126/sciadv.abg2616},
	number = {33},
	urldate = {2025-07-21},
	journal = {Sci. Adv.},
	author = {Hänze, Max and McMurtrie, Gregory and Baumann, Susanne and Malavolti, Luigi and Coppersmith, Susan N. and Loth, Sebastian},
	month = aug,
	year = {2021},
}

@article{Mari2015,
doi = {10.1088/0953-4075/48/17/175501},
url = {https://dx.doi.org/10.1088/0953-4075/48/17/175501},
year = {2015},
month = {jul},
publisher = {IOP Publishing},
volume = {48},
number = {17},
pages = {175501},
author = {Mari, A and Farace, A and Giovannetti, V},
title = {Quantum optomechanical piston engines powered by heat},
journal = {J. Phys. B: At. Mol. Opt. Phys.},
}

@article{SerraGarica2016,
  title = {Mechanical Autonomous Stochastic Heat Engine},
  author = {Serra-Garcia, Marc and Foehr, Andr\'e and Moler\'on, Miguel and Lydon, Joseph and Chong, Christopher and Daraio, Chiara},
  journal = {Phys. Rev. Lett.},
  volume = {117},
  issue = {1},
  pages = {010602},
  numpages = {6},
  year = {2016},
  month = {Jun},
  publisher = {American Physical Society},
  doi = {10.1103/PhysRevLett.117.010602},
  url = {https://link.aps.org/doi/10.1103/PhysRevLett.117.010602}
}

@article{Carollo2020,
  title = {Nonequilibrium Many-Body Quantum Engine Driven by Time-Translation Symmetry Breaking},
  author = {Carollo, Federico and Brandner, Kay and Lesanovsky, Igor},
  journal = {Phys. Rev. Lett.},
  volume = {125},
  issue = {24},
  pages = {240602},
  numpages = {6},
  year = {2020},
  month = {Dec},
  publisher = {American Physical Society},
  doi = {10.1103/PhysRevLett.125.240602},
  url = {https://link.aps.org/doi/10.1103/PhysRevLett.125.240602}
}

@article{Guzman2024,
doi = {10.1088/1361-6633/ad8803},
url = {https://dx.doi.org/10.1088/1361-6633/ad8803},
year = {2024},
month = {nov},
publisher = {IOP Publishing},
volume = {87},
number = {12},
pages = {122001},
author = {Antonio Marín Guzmán, José and Erker, Paul and Gasparinetti, Simone and Huber, Marcus and Yunger Halpern, Nicole},
title = {Key issues review: useful autonomous quantum machines},
journal = {Rep. Prog. Phys.},
}

@ARTICLE{Frunzio2005,
  author={Frunzio, L. and Wallraff, A. and Schuster, D. and Majer, J. and Schoelkopf, R.},
  journal={IEEE Trans. Appl. Supercond.}, 
  title={Fabrication and characterization of superconducting circuit QED devices for quantum computation}, 
  year={2005},
  volume={15},
  number={2},
  pages={860-863},
  doi={10.1109/TASC.2005.850084}
}

@article{Frattini2017,
    author = {Frattini, N. E. and Vool, U. and Shankar, S. and Narla, A. and Sliwa, K. M. and Devoret, M. H.},
    title = {3-wave mixing Josephson dipole element},
    journal = {Appl. Phys. Lett.},
    volume = {110},
    number = {22},
    pages = {222603},
    year = {2017},
    month = {05},
    issn = {0003-6951},
    doi = {10.1063/1.4984142},
    url = {https://doi.org/10.1063/1.4984142},
}

@article{Reagor2013,
    author = {Reagor, Matthew and Paik, Hanhee and Catelani, Gianluigi and Sun, Luyan and Axline, Christopher and Holland, Eric and Pop, Ioan M. and Masluk, Nicholas A. and Brecht, Teresa and Frunzio, Luigi and Devoret, Michel H. and Glazman, Leonid and Schoelkopf, Robert J.},
    title = {Reaching 10 ms single photon lifetimes for superconducting aluminum cavities},
    journal = {Appl. Phys. Lett.},
    volume = {102},
    number = {19},
    pages = {192604},
    year = {2013},
    month = {05},
    issn = {0003-6951},
    doi = {10.1063/1.4807015},
    url = {https://doi.org/10.1063/1.4807015},
}

@book{GardinerZoller2004,
  title={Quantum Noise},
  author={Gardiner, Crispin and Zoller, Peter},
  year={2004},
  publisher={Springer}
}

@article{Somoroff2023,
  title = {Millisecond Coherence in a Superconducting Qubit},
  author = {Somoroff, Aaron and Ficheux, Quentin and Mencia, Raymond A. and Xiong, Haonan and Kuzmin, Roman and Manucharyan, Vladimir E.},
  journal = {Phys. Rev. Lett.},
  volume = {130},
  issue = {26},
  pages = {267001},
  numpages = {6},
  year = {2023},
  month = {Jun},
  publisher = {American Physical Society},
  doi = {10.1103/PhysRevLett.130.267001},
  url = {https://link.aps.org/doi/10.1103/PhysRevLett.130.267001}
}

@article{Gillespie1977,
 author = {Gillespie, Daniel},
 title = {Exact {stochastic} {simulation} of {coupled} chemical reactions},
 journal = {J. Phys. Chem.},
 volume = {81},
 number = {25},
 pages = {2340},
 year = {1977},
 doi = {10.1021/j100540a008},
 url = {https://pubs.acs.org/doi/abs/10.1021/j100540a008}
}

\end{document}